\documentclass[%
superscriptaddress,
reprint, aps,prl,
 amsmath,amssymb,
floatfix,nofootinbib
]{revtex4-1}

\usepackage{csquotes}
\usepackage{graphicx}
\usepackage{dcolumn}
\usepackage{bm}
\usepackage[usenames,dvipsnames,svgnames,table]{xcolor}
\usepackage{selinput}
\usepackage{soul}
\usepackage[normalem]{ulem}
\usepackage{cancel}
\usepackage{sansmathfonts}
\usepackage{amsmath}
\usepackage{txfonts}
\usepackage{multirow}
\usepackage[T1]{fontenc}

\graphicspath{{./}}

\begin{document}

\title{Beyond the Tip: Lattice Dynamics, Seams, and the Mechanism of Microtubule Fracture}

\author{Amir Zablotsky}
\affiliation{Université Grenoble-Alpes, CNRS, Laboratoire Interdisciplinaire de Physique 38000 Grenoble, France}

\author{Subham Biswas}
\affiliation{Experimental Physics and Center for Biophysics, Saarland University, 66123 Saarbrücken, Germany}

\author{Laura Schaedel}
\email{laura.schaedel@uni-saarland.de}
\affiliation{Experimental Physics and Center for Biophysics, Saarland University, 66123 Saarbrücken, Germany}
\affiliation{PharmaScienceHub (PSH), 66123 Saarbrücken, Germany}

\author{Karin John}
\email{karin.john@univ-grenoble-alpes.fr}
\affiliation{Université Grenoble-Alpes, CNRS, Laboratoire Interdisciplinaire de Physique 38000 Grenoble, France}


\date{\today}

\begin{abstract}
The structural integrity of microtubules is paramount for cellular function. We present a theoretical analysis of their lattice fracture, focusing on the influence of multi-seam structures arising from monomer defects and aiming to provide a more accurate estimation of GDP lattice parameters. Our findings reveal that seams function as pre-existing pathways that accelerate damage propagation. Consequently,  monomer vacancies destabilize the lattice due to the inherent structural loss of tubulin-tubulin contacts and the additive acceleration of fracture through multiple seams. Importantly, the comparison of our simulations with experiments on lattice fracture suggests that the intrinsic ratio of longitudinal to lateral binding energies is bounded at approximately 1.5, challenging previous predictions of lattice anisotropy from tip-growth models and highlighting the urgent need to incorporate into current growth models parameters obtained from lattice dynamics.
\end{abstract}

\maketitle


Microtubules (MTs) are central structures in living cells, involved in cell division, migration, and intracellular transport \cite{AlbertsBook}. A complete understanding of the mechanisms regulating their dynamics and stability is a central issue in cell biology \cite{Gudimchuk2021} and a key challenge for human health \cite{Sferra2020,Wordeman2021}.
Common textbook knowledge states that microtubules are  hollow cylindrical structures formed by $\alpha\beta-$tubulin heterodimers arranged in a quasi-crystalline lattice, whose dynamics is restricted to elongation and shortening at their tips. 
Upon incorporation into the lattice, GTP-bound tubulin undergoes hydrolysis to GDP, transforming the stable GTP lattice at the growing plus end into a conformationally strained and less stable GDP lattice in the bulk. This irreversible hydrolysis-driven decrease in lattice stability is at the origin of stochastic switches between phases of growth and  rapid depolymerization.
Although the discovery of this dynamic instability \cite{Mitchison1984} has spurred extensive research into MT tip dynamics \cite{Akhmanova2015,Gudimchuk2021,Cleary2021}, the dynamics of the bulk lattice has received less attention. This is largely due to the prevailing notion that it exists in a frozen, static state and the historical lack of techniques to investigate the meta-stable GDP lattice. 

Recent experiments utilizing end-stabilized microtubules (MTs) have revealed lattice dynamics and self-repair \cite{Dye1992,Schaedel2015,Aumeier2016,Vemu2018,Schaedel2019,Triclin2021,Alexandrova2022,AndreuCarbo2022,Budaitis2022,Gazzola2023,Biswas2025} distant from the tip, revitalizing interest in bulk lattice properties. This renewed focus is further fueled by the observation of MT fracture in vivo, whether induced by severing enzymes  \cite{McNally2018} or mechanical forces \cite{TangSchomer2009,Nandakumar2025}, which has spurred investigation into the underlying fracture pathway.
\begin{figure}[hbt]
\centering
\includegraphics[width=0.9\hsize]{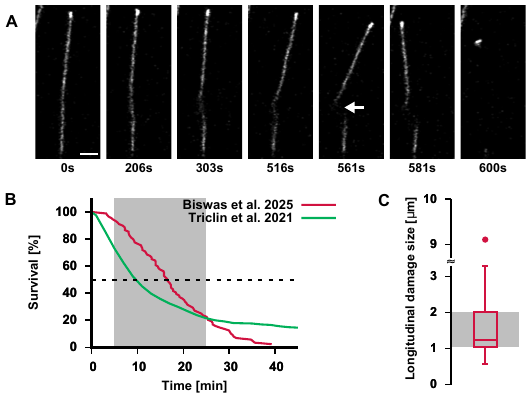}
\caption{
\textbf{A:} Image sequence showing a MT developing a damaged region, which is visible due to reduced fluorescence resulting from the loss of tubulin from the lattice. The MT eventually broke along the softened region (marked by the white arrow) and disassembled. Scale bar, $2\,\muup$m. See Ref.\,\cite{Biswas2025} for experimental details.
\textbf{B:} Survival times of end-stabilized microtubules (experimental data from Refs.\,\cite{Triclin2021,Biswas2025}). 
\textbf{C:} Size of damage at fracture (experimental data from Ref.\,\cite{Biswas2025}). Shown is the median and the interquartile range (IQR).
The gray shaded regions in (B,C) indicate the range of admissible values we chose for comparison with our simulation results.
}
\label{fig-model_fracture}
\end{figure}
Even in the absence of external forces and cellular factors (e.g. severing enzymes) and without free tubulin, end-stabilized microtubules begin to depolymerize from within the lattice (Fig.\,\ref{fig-model_fracture}A) before eventually breaking into two distinct fragments. These new ends then undergo rapid depolymerization. The time to fracture is typically between 10 and 20\,minutes  (Fig.\,\ref{fig-model_fracture}B, \cite{Schaedel2019,Triclin2021,Biswas2025}), at which point the damaged lattice region, marked by a diminished fluorescence, spans on average about 1\,$\muup$m along the MT axis (Fig.\,\ref{fig-model_fracture}C). 

The experimental quantification of dimer loss from the lattice 
in combination with a kinetic lattice model offers a powerful approach to study the path to fracture from an initial defect and to deduce critical GDP lattice parameters. 
Our study therefore leverages this approach to investigate microtubule lattice dynamics in the absence of free tubulin, specifically examining how the location and nature of an initial defect influences microtubule fracture.
While our earlier work  \cite{Biswas2025} identified the MT fracture as a readout for lattice anisotropy, we now qualitatively and quantitatively dissect this process, as well as the role of seam structures and defects in driving fracture dynamics. Specifically, we ask, how the lattice geometry determines the path to fracture.
Furthermore, we identify the lattice parameters that best reproduce experimental MT fracture data and compare these parameters to those found in the existing literature.
Our study shows that monomer vacancies, the most abundant topological defects in MTs \cite{Guyomar2022} significantly impact MT fracture and that the lattice anisotropy is much weaker than previously estimated from e.g.~tip growth models, requiring a revision of current lattice models.\\


\begin{figure}[hbt]
\centering
\includegraphics[width=0.97\hsize]{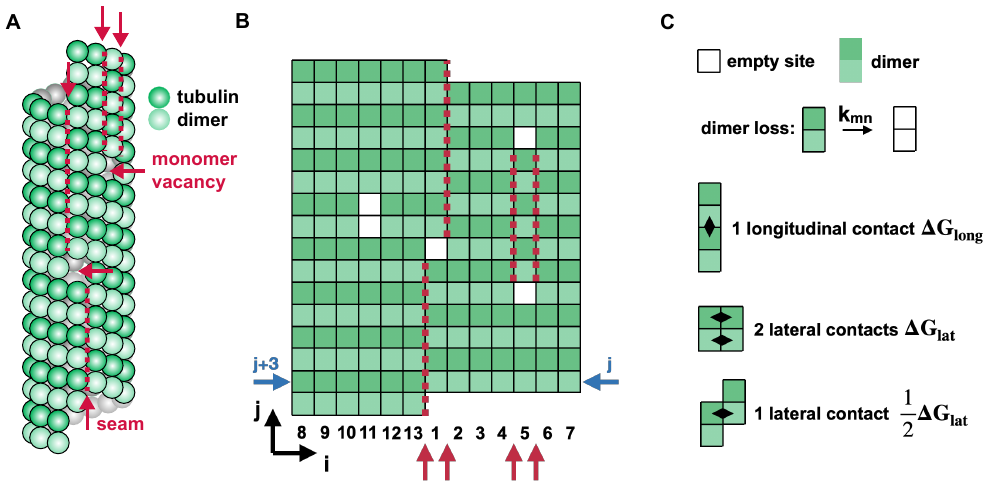}
\caption{
\textbf{A:} Three dimensional model of the canonical microtubule structure (13 protofilament, 3-start helix). Monomer vacancies (horizontal red arrows) lead either to a lateral shift of an existing seam or introduce two new seams. Seam positions are marked by vertical red arrows and red dashed lines. 
\textbf{B:} Schematic of the lattice setup with seam structures shown in red. The transition between protofilaments $i=13$ and $i=1$ corresponds to a periodic boundary with a vertical shift of 3 monomers (as indicated by the blue annotated arrows) to assure the 3-start helical lattice structure.
\textbf{C:} The dimer off-rate constant $k_{mn}$ (see Eq.~\eqref{ec-detachment_rates}) depends on the occupancy of neighboring lattice sites. Interactions between neighboring longitudinal or lateral occupied lattice sites (diamond symbols) are characterized by the binding energies $\Delta G_\text{long}$ and $\Delta G_\text{lat}$, respectively.} 
\label{fig:model}
\end{figure}

We implement a kinetic Monte Carlo model (\cite{Lukkien1998} and Supplemental Material \cite{SupMat}), which is frequently used to study MT dynamics \cite{vanBuren2002,Gardner2011, Margolin2012,Piedra2016,Chaaban2018,Mickolajczyk2019,Thawani2020,McCormick2024}. Briefly, the canonical MT structure (Fig.\,\ref{fig:model}A) is represented as a two-dimensional lattice at the monomer scale (Fig.\,\ref{fig:model}B). 
The lattice may contain monomer vacancies. If such a monomer vacancy is directly adjacent to a seam, the seam is shifted laterally by one protofilament; if the monomer vacancy is in the bulk lattice, two new seams will emanate from the vacancy as depicted in Fig.\,\ref{fig:model}A,B.
Each dimer interacts with up to four lateral monomer neighbors (binding energy ${1\over 2}\Delta G_\text{lat}$ per lateral monomer-monomer contact) and with up to two longitudinal neighbors (binding energy $\Delta G_\text{long}$ per longitudinal contact) resulting in the total binding energy for a fully surrounded dimer of
\begin{equation}
    \Delta G_\text{b}=2\Delta G_\text{long}+2\Delta G_\text{lat}\,.
    \label{ec-binding_energy}
\end{equation}
The lattice anisotropy determines how strong the longitudinal interactions are compared to the lateral ones and is defined here as\footnote{We favor here the measure \eqref{ec-lattice_anisotropy} over a more classical definition of the anisotropy  $(\Delta G_\mathrm{long}-\Delta G_\mathrm{lat})/(\Delta G_\mathrm{long}+\Delta G_\mathrm{lat})$ since it facilitates a direct comparison with data in the literature.} 
\begin{equation}
A=\frac{\Delta G_\text{long}}{\Delta G_\text{lat}}\,.
\label{ec-lattice_anisotropy}
\end{equation}
In the absence of free tubulin the only transition relevant in the GDP lattice is the detachment of dimers (rate constant $k_\text{mn}$), which depends on the number of longitudinal ($m$) and lateral ($n$) monomer contacts via the Arrhenius equation \cite{Kondepudi2014}
\begin{equation}
    k_{mn}=\frac{1}{\tau} e^{\beta\left(m\Delta G_\text{long}+{n\over 2}\Delta G_\text{lat}-{\Delta G_\text{b}\over 2}\right)}\,.
\label{ec-detachment_rates}
\end{equation}
$\beta$ denotes the inverse of the thermodynamic temperature. $1/\tau$ is the off-rate constant of a corner dimer with $m=1$ longitudinal and $n=2$ lateral monomer neighbors. Dimer detachment and reattachment at nearby vacant lattice sites are not considered.
In our simulations we study the dynamics of 10\,$\muup$m long MTs stabilized at both ends by a non-detachable seed/cap structure, comparable to experiments \cite{Schaedel2015,Schaedel2019,Triclin2021,Biswas2025}.\\  


\begin{figure*}[hbt]
\centering
\includegraphics[width=0.9\hsize]{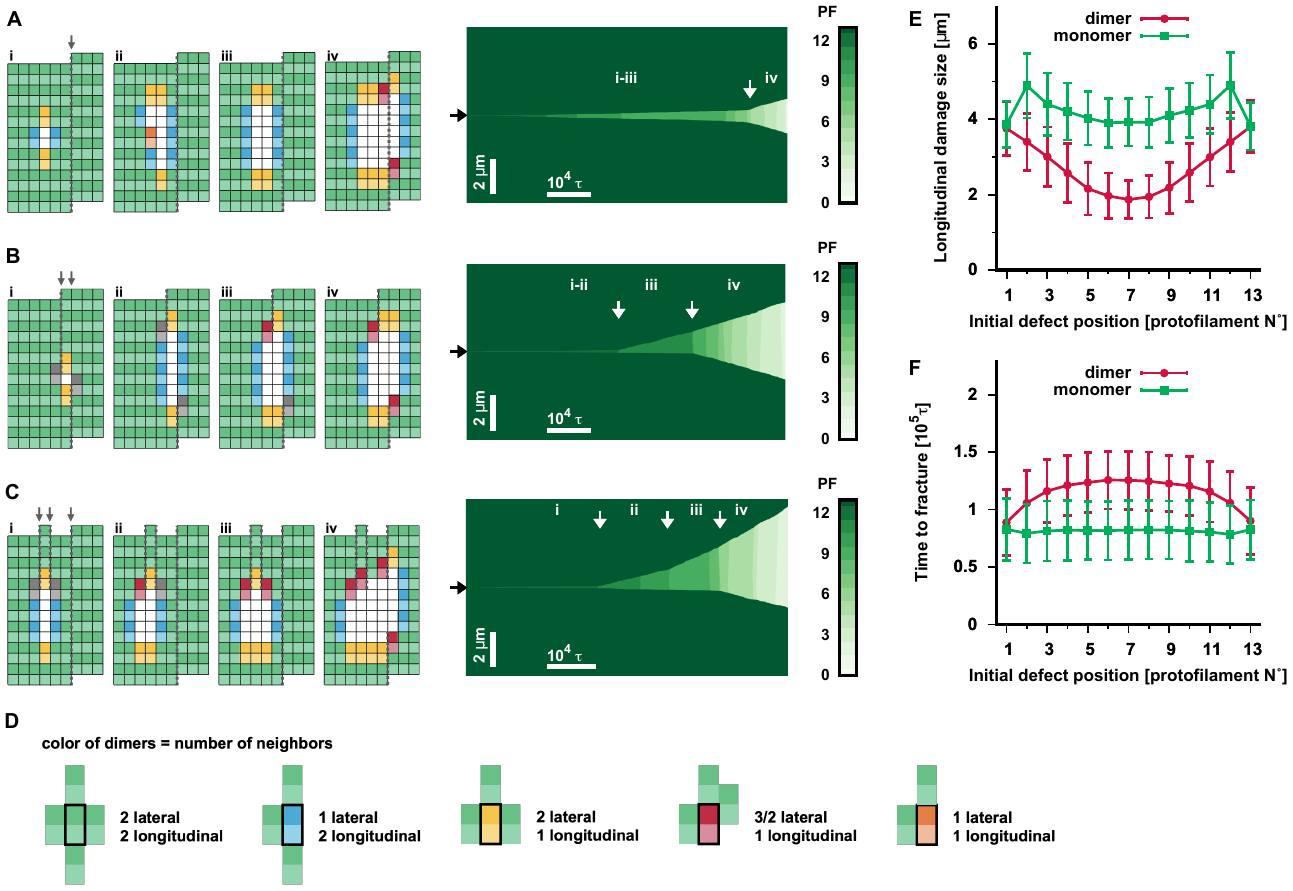}
\caption{Path to fracture in multi-seam microtubules.
\textbf{A-D:} Vacancy growth from a dimer vacancy (A) and a monomer vacancy situated at an existing seam (B) or in the B-lattice (C). Dimer color codes (D): most stable (left) to least stable (right). \textit{Left:} Illustrations of the lattice configurations in the vicinity of the growing vacancy at different propagation stages of vacancy growth. The colors distinguish the dimers by number of longitudinal and lateral neighbors.  \textit{Right:} Exemplary kymographs of the fracture process. The transition of seam structures are marked by white arrows. The $y-$axis represents the longitudinal axis of the microtubule and the color scale indicates the number of intact protofilaments. 
\textbf{E,F: } Examples of the longitudinal damage size at fracture (E) and time to fracture (F; in units of $\tau$) as a function of the initial lateral position (in protofilaments) of the initial defect. Symbols and error bars indicate the mean and standard deviation of the sample (1000 realizations).  
Parameters are $\Delta G_\mathrm{b}=-45\,$kT and $A=1.7$.}
\label{fig:path}
\end{figure*}

First we investigate tubulin loss from the lattice upon placing a dimer or monomer vacancy at a given position in the lattice. Thereby we concentrate on lattice energies $-\Delta G_\text{b}\gg\text{kT}$, such that fracture will proceed predominantly from the initial vacancy and spontaneous dimer loss from the fully occupied lattice is a rare event on this time scale. 
The typical paths to fracture are shown in Fig.\,\ref{fig:path}A-C (left).
For the experimentally relevant case with anisotropy $A>1$, the vacancy (config.\,(i) in Fig.\,\ref{fig:path}A-C) will grow by loosing dimers primarily in the longitudinal direction, since $k_{14}>k_{22}$. A dimer loss in the lateral direction produces a short lived configuration with a corner dimer (orange in config.\,(ii) in Fig.\,\ref{fig:path}A). Since the detachment of corner dimers is much more rapid than the detachment of other boundary dimers (blue and yellow in Fig.\,\ref{fig:path}A, $k_{12}=1/\tau\gg k_{14} > k_{22}$), the vacancy will essentially grow with a rectangular shape (config.\,(iii) in Fig.\,\ref{fig:path}A). 
The longitudinal (lateral) propagation speed 
can be approximated by $v_\mathrm{long}=2\eta N_\mathrm{lat} k_{14}$ ($v_\mathrm{lat}=\eta N_\mathrm{long}k_{22}$)
where $N_\mathrm{lat}$ ($N_\mathrm{long}$) denotes the lateral (longitudinal) extension of the damage and  $\eta=4\,$nm denotes the size of a monomer.
 As the seam is crossed in the lateral direction, the longitudinal boundaries will contain a seam dimer (red in config.\,(iv) in Fig.\,\ref{fig:path}A-C) with one longitudinal and three lateral monomer neighbors, which increases the longitudinal front speed to
$v^\mathrm{seam}_\mathrm{long} \ge \eta k_{13}$. Note, that this acceleration is a structural effect and not related to an energetic difference between heterotypic and homotypic lateral contacts, which is not considered here. The estimate for $v^\mathrm{seam}_\mathrm{long}$ is valid for anisotropies $A<2.5$. For higher lattice anisotropies $(1,4)$-dimers also contribute to the front velocity. MT fracture is complete when the damage spans all 13 protofilaments. 

When a monomer vacancy is initially positioned next to a seam, the first lateral detachment event triggers an immediate seam crossing for one  longitudinal front (up in config.\,(iii) in Fig.\,\ref{fig:path}B). 
This process disrupts the symmetry between the upward- and downward-moving fronts, as only one front gains a seam dimer, thereby increasing its velocity.
Once the vacancy traverses the second seam, both longitudinal fronts propagate with the same velocity (config.~(iv) in Fig.\,\ref{fig:path}B) reestablishing symmetry in the front propagation. 
For a monomer defect initially placed within the B-lattice, which leads to the formation of two additional seams extending towards one side of the vacancy (Fig.\,\ref{fig:path}C), the lateral growth of the vacancy over each seam contributes one additional seam dimer to the longitudinal boundary that traverses the seam. Most importantly, this results in an asymmetric front propagation towards the two microtubule extremities, with the front propagating faster in the direction of the multi seam structure than towards the perfect B-lattice (config.\,(iii) in Fig.\,\ref{fig:path}C). In more general terms, the longitudinal front speed increases with the number of seams that originate from the front (config.\,(iv) in Fig.\,\ref{fig:path}C). 

The kymographs in Fig.\,\ref{fig:path}A-C (right) visualize how an initial vacancy propagates through the lattice until complete microtubule breakage. The longitudinal front speed increases significantly, when the vacancy crosses a seam (white arrows in Fig.\,\ref{fig:path}A-C).
Note, that the asymmetry in the longitudinal front speed is completely absent for vacancy growth in the perfect lattice (Fig.\,\ref{fig:path}A) and presents a hallmark of multi seam structures.

The structural role of seam dimers in vacancy growth is reflected in two measurable properties: the longitudinal damage size at fracture ($L_\text{f}$, Fig.\,\ref{fig:path}E) and the time to fracture ($T_\text{f}$, Fig.~\ref{fig:path}F). For dimer vacancies, $L_\text{f}$ is maximal when the defect originates near the seam (protofilament 1 or 13) and decreases (approx. 2-fold for the parameters in Fig.\,\ref{fig:path}E) as the initial vacancy moves away from it. Conversely, $T_\text{f}$ is shortest at the seam and increases with distance. Monomer defects produce consistently longer $L_\text{f}$ and shorter $T_f$ than dimer defects without significant dependence on the defect's initial position.
This distinct behavior is because a monomer defect, independent of its initial position, requires only a single lateral detachment to create a seam dimer, after which rapid longitudinal propagation governs the path to breakage.
Note, that the broad distributions of $L_\text{f}$ and $T_\text{f}$ arise (i) from the stochastic choice of pathways of vacancy growth and (ii) the stochasticity of dimer detachment following an exponential distribution.\\

\begin{figure}[hbt]
\centering
\includegraphics[width=1\hsize]{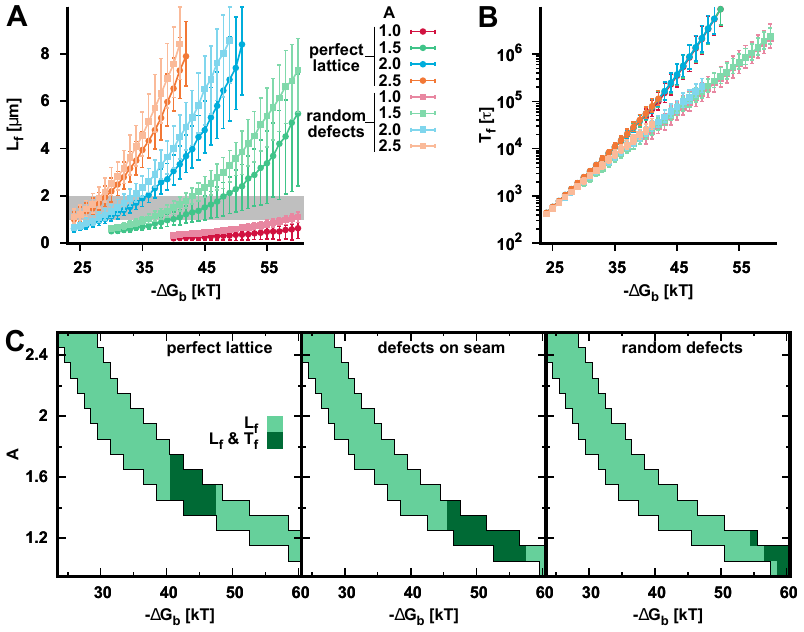}
\caption{
Fracture damage size ($L_\text{f}$, A) and time ($T_\text{f}$, B) vs. lattice binding energy ($\Delta G_\text{b}$) for varying anisotropies ($A$). Results for defect-free lattices and uniformly distributed defects; seam-localized defects yield similar trends (Fig.\,S1\,A,B). Data: median $\pm$ IQR (1000 simulations). Gray region in (A): experimental range (Fig.\,\ref{fig-model_fracture}C).
\textbf{C: } Lattice parameters $\left(\Delta G_b,A\right)$ that reproduce experiments assuming defect free MTs (perfect lattice) and MTs containing defects localized on the main seam (defects on seam) or distributed uniformly throughout the lattice (random defects).
Light green: parameters reproducing $L_\text{f}$; dark green: parameters reproducing both $L_\text{f}$ and $T_\text{f}$.
The spatial frequency of defects in (A-C) is 0.2\,$\muup$m$^{-1}$. Further results for a frequency of 0.1\,$\muup$m$^{-1}$ are shown in Fig.\,S2 in Supplemental Material \cite{SupMat}.
}
\label{fig:exp}
\end{figure}

To link our model with experiments (Fig.\,\ref{fig-model_fracture}B,C), we simulated the disassembly of 10\,$\muup$m end-stabilized microtubules, comparing a perfect canonical lattice\footnote{Note, that in the perfect lattice the first vacancy is created by the spontaneous loss of a dimer with 2 longitudinal and 4 lateral monomer neighbors.}  with a lattice containing random monomer vacancies at a realistic frequency (0.2\,$\muup$m$^{-1}$  \cite{Biswas2025}). We examined both uniformly distributed defects and defects localized at the main seam.
We calculated the fracture damage size $L_{\text{f}}$ and the survival time $T_{\text{f}}$ for various combinations of $\Delta G_\text{b}$ and $A$ (Fig.\,\ref{fig:exp}A,B and Fig.\,S1 in Supplemental Material \cite{SupMat}). 
$L_\text{f}$ and $T_\text{f}$ both increase with $-\Delta G_\text{b}$. $L_\text{f}$ increases with $A$, while $T_\text{f}$ only weakly depends on $A$ (these features are further discussed in Supplemental Material \cite{SupMat}). 
To set a relevant timescale $\tau$ for $T_\text{f}$, we calibrated our model to match experimental values for plus-end tip depolymerization speed ($v = 10 - 30\text{ }\muup\text{m}.\text{min}^{-1}$, \cite{Mitchison1984,Cassimeris1986,Walker1988,Verde1992,Luchniak2023}, see Supplemental Material \cite{SupMat} for details).
The shaded regions in Fig.\,\ref{fig:exp}C highlight the lattice parameters which reproduce $L_\text{f}$. Specifically, the dark green regions mark the parameter space where both $L_{\text{f}}$ and $T_{\text{f}}$ are reproduced. 
This process yielded key parameter estimates: a perfect canonical lattice suggests a total binding energy $\Delta G_{\text{b}} \sim -45\,$kT and a lattice anisotropy $A \sim 1.5$. The presence of defects implies higher binding energy and lower lattice anisotropy, depending on the spatial distribution of defects within the lattice.

Comparison of these GDP lattice parameters with literature values from kinetic Monte Carlo (KMC) simulations of microtubule tip dynamics (\cite{vanBuren2002,Margolin2012,Piedra2016,Chaaban2018,McCormick2024} and Supplemental Material \cite{SupMat}) reveals close agreement in binding energy: $\Delta G_{\text{b}}$ typically ranges from -35 to -55\,kT in literature, aligning well with our derived range of -45 to -50\,kT. However, a discrepancy exists regarding anisotropy ($A$). While Refs.~\cite{vanBuren2002, Margolin2012,Chaaban2018} reported $A > 3$, the studies specifically investigating GDP tubulin's effect on tip dynamics (Refs.~\cite{Piedra2016,McCormick2024}) found values ($A = 1.8$ and $A \approx 1.4$) that approximately match our own findings for the perfect lattice.

Estimates derived from all-atom molecular dynamics (MD) simulations, which explicitly quantify the binding energy of the GDP-lattice \cite{Sept2003,Kononova2014}, typically yield higher binding energies than those reported by KMC tip models. When these MD-derived parameters are used in our model ($\Delta G_{\text{b}} \approx -110$\,kT, $A = 1.5$ \cite{Sept2003}; $\Delta G_{\text{b}} \approx -76\,$kT, $A = 2.2$  \cite{Kononova2014}), they result in predicted damage sizes at fracture significantly exceeding $10\,\muup\text{m}$.
A combination of all-atom molecular dynamics and brownian dynamics simulations \cite{Hemmat2019} for the tip growth yields similar results as the KMC models with binding energy comparable to our model but a high lattice anisotropy ($\Delta G_\text{b}=-37\,$kT, $A=2.9$, Supplemental Material \cite{SupMat}).

A third class, so called chemo-mechanical models  \cite{vanBuren2005,Coombes2013,Zakharov2015,McIntosh2018,Gudimchuk2020}, combine kinetic Monte Carlo models with either bead-spring mechanical models or coarse-grained molecular dynamics simulations (monomer scale). Ref.~\cite{Gudimchuk2020} explicitly tabulates values for the GDP-lattice (neglecting the lattice strain energy) with $\Delta G_\text{b}=-65\,$kT and $A=1.1$, which would match the size of damage but not the time of fracture in our model.   

Finally, the analysis of experimental force-indentation curves using high-speed atomic force microscopy on taxolated microtubules \cite{Ganser2019} estimates a total binding energy of the order $\Delta G_\text{b}=-60\,$kT with an anisotropy $A\approx 5$, suggesting also a very rapid propagation of lattice damage in the longitudinal direction, causing far bigger damage sizes at fracture than observed experimentally in the unstabilized GDP-lattice.\\ 


In conclusion, our qualitative and quantitative analysis of microtubule fracture dynamics reveals two key findings.
First, damage is driven by two structural features: the seam lattice geometry, which accelerates longitudinal vacancy propagation compared to the B-lattice, and the presence of monomer defects. These defects impact the lattice in three critical ways: by inherently weakening it through contact loss, by breaking the symmetry of damage propagation (otherwise symmetrical), and by forming multi-seam structures that amplify longitudinal progression. Given their high prevalence \cite{Guyomar2022} compared to other defect types (e.g., changes in protofilament number) and their ability to alter the microtubule seam structure, monomer vacancies emerge as a key origin of microtubule fracture. Second, previous estimations may have overestimated the GDP-lattice anisotropy; our simulations suggest a lower anisotropy (ratio of longitudinal to lateral binding energies) of about $A = 1.5$, a value further reduced by even rare monomer vacancies.

Estimating lattice parameters from microtubule tip growth alone is challenging due to the combined influence of GTP and GDP-lattice properties. However, studying microtubule fracture and the role of monomer defects provides an additional means of accessing GDP-lattice properties.
Here, we used a minimal model for the GDP lattice, where homotypic and heterotypic lateral contacts were treated as identical. However, it is straightforward to extend our model and include a weak destabilization of seam structures \cite{Katsuki2014}, which would most likely decrease the estimate of the lattice anisotropy.
Furthermore, we do not consider coupling between lattice stress/strain and lattice dynamics; ideally a mechano-chemical lattice model similar to models developed for tip growth \cite{Zakharov2015,McIntosh2018,Gudimchuk2020} is needed to understand MT fracture in more detail.  
However, our findings underscore a critical limitation in current growth models:  they do not account for the documented presence of monomer defects. A critical revision of tip growth that provides a plausible mechanism of vacancy formation is essential for a comprehensive model of microtubule assembly.  

\begin{acknowledgements}
A.Z. and K.J. were supported through the Idex Université Grenoble-Alpes. The numerical computations by A.Z. and K.J. were performed using the Cactus cluster of the CIMENT infrastructure, which was supported by the Rhône-Alpes region. We are grateful to P. Beys, who manages the cluster.
S.B. and L.S. were supported by the Deutsche Forschungsgemeinschaft (DFG, German Research Foundation; grant number SFB 1027/A13) and through Saarland University’s NanoBioMed Young Investigator Grant awarded to L.S.
\end{acknowledgements}

\paragraph{Data availability:} The data that support the findings of this article are openly available \cite{Zablotsky2026Zenodo}.


\begin{thebibliography}{48}%
\makeatletter
\providecommand \@ifxundefined [1]{%
 \@ifx{#1\undefined}
}%
\providecommand \@ifnum [1]{%
 \ifnum #1\expandafter \@firstoftwo
 \else \expandafter \@secondoftwo
 \fi
}%
\providecommand \@ifx [1]{%
 \ifx #1\expandafter \@firstoftwo
 \else \expandafter \@secondoftwo
 \fi
}%
\providecommand \natexlab [1]{#1}%
\providecommand \enquote  [1]{``#1''}%
\providecommand \bibnamefont  [1]{#1}%
\providecommand \bibfnamefont [1]{#1}%
\providecommand \citenamefont [1]{#1}%
\providecommand \href@noop [0]{\@secondoftwo}%
\providecommand \href [0]{\begingroup \@sanitize@url \@href}%
\providecommand \@href[1]{\@@startlink{#1}\@@href}%
\providecommand \@@href[1]{\endgroup#1\@@endlink}%
\providecommand \@sanitize@url [0]{\catcode `\\12\catcode `\$12\catcode
  `\&12\catcode `\#12\catcode `\^12\catcode `\_12\catcode `\%12\relax}%
\providecommand \@@startlink[1]{}%
\providecommand \@@endlink[0]{}%
\providecommand \url  [0]{\begingroup\@sanitize@url \@url }%
\providecommand \@url [1]{\endgroup\@href {#1}{\urlprefix }}%
\providecommand \urlprefix  [0]{URL }%
\providecommand \Eprint [0]{\href }%
\providecommand \doibase [0]{http://dx.doi.org/}%
\providecommand \selectlanguage [0]{\@gobble}%
\providecommand \bibinfo  [0]{\@secondoftwo}%
\providecommand \bibfield  [0]{\@secondoftwo}%
\providecommand \translation [1]{[#1]}%
\providecommand \BibitemOpen [0]{}%
\providecommand \bibitemStop [0]{}%
\providecommand \bibitemNoStop [0]{.\EOS\space}%
\providecommand \EOS [0]{\spacefactor3000\relax}%
\providecommand \BibitemShut  [1]{\csname bibitem#1\endcsname}%
\let\auto@bib@innerbib\@empty
\bibitem [{\citenamefont {Alberts}\ \emph {et~al.}(2022)\citenamefont
  {Alberts}, \citenamefont {Heald}, \citenamefont {Johnson}, \citenamefont
  {Morgan}, \citenamefont {Raff}, \citenamefont {Roberts},\ and\ \citenamefont
  {Walter}}]{AlbertsBook}%
  \BibitemOpen
  \bibfield  {author} {\bibinfo {author} {\bibfnamefont {B.}~\bibnamefont
  {Alberts}}, \bibinfo {author} {\bibfnamefont {R.}~\bibnamefont {Heald}},
  \bibinfo {author} {\bibfnamefont {A.}~\bibnamefont {Johnson}}, \bibinfo
  {author} {\bibfnamefont {D.}~\bibnamefont {Morgan}}, \bibinfo {author}
  {\bibfnamefont {M.}~\bibnamefont {Raff}}, \bibinfo {author} {\bibfnamefont
  {K.}~\bibnamefont {Roberts}}, \ and\ \bibinfo {author} {\bibfnamefont
  {P.}~\bibnamefont {Walter}},\ }\href@noop {} {\emph {\bibinfo {title}
  {Molecular Biology of the Cell}}}\ (\bibinfo  {publisher} {WW Norton \&
  Company},\ \bibinfo {year} {2022})\BibitemShut {NoStop}%
\bibitem [{\citenamefont {Gudimchuk}\ and\ \citenamefont
  {McIntosh}(2021)}]{Gudimchuk2021}%
  \BibitemOpen
  \bibfield  {author} {\bibinfo {author} {\bibfnamefont {N.~B.}\ \bibnamefont
  {Gudimchuk}}\ and\ \bibinfo {author} {\bibfnamefont {J.~R.}\ \bibnamefont
  {McIntosh}},\ }\href {\doibase 10.1038/s41580-021-00399-x} {\bibfield
  {journal} {\bibinfo  {journal} {Nat. Rev. Mol. Cell Biol.}\ }\textbf
  {\bibinfo {volume} {22}},\ \bibinfo {pages} {777} (\bibinfo {year}
  {2021})}\BibitemShut {NoStop}%
\bibitem [{\citenamefont {Sferra}\ \emph {et~al.}(2020)\citenamefont {Sferra},
  \citenamefont {Nicita},\ and\ \citenamefont {Bertini}}]{Sferra2020}%
  \BibitemOpen
  \bibfield  {author} {\bibinfo {author} {\bibfnamefont {A.}~\bibnamefont
  {Sferra}}, \bibinfo {author} {\bibfnamefont {F.}~\bibnamefont {Nicita}}, \
  and\ \bibinfo {author} {\bibfnamefont {E.}~\bibnamefont {Bertini}},\ }\href
  {\doibase 10.3390/ijms21197354} {\bibfield  {journal} {\bibinfo  {journal}
  {Int. J. Mol. Sci.}\ }\textbf {\bibinfo {volume} {21}},\ \bibinfo {pages}
  {7354} (\bibinfo {year} {2020})}\BibitemShut {NoStop}%
\bibitem [{\citenamefont {Wordeman}\ and\ \citenamefont
  {Vicente}(2021)}]{Wordeman2021}%
  \BibitemOpen
  \bibfield  {author} {\bibinfo {author} {\bibfnamefont {L.}~\bibnamefont
  {Wordeman}}\ and\ \bibinfo {author} {\bibfnamefont {J.~J.}\ \bibnamefont
  {Vicente}},\ }\href {\doibase 10.3390/cancers13225650} {\bibfield  {journal}
  {\bibinfo  {journal} {Cancers}\ }\textbf {\bibinfo {volume} {13}},\ \bibinfo
  {pages} {5650} (\bibinfo {year} {2021})}\BibitemShut {NoStop}%
\bibitem [{\citenamefont {Mitchison}\ and\ \citenamefont
  {Kirschner}(1984)}]{Mitchison1984}%
  \BibitemOpen
  \bibfield  {author} {\bibinfo {author} {\bibfnamefont {T.}~\bibnamefont
  {Mitchison}}\ and\ \bibinfo {author} {\bibfnamefont {M.~W.}\ \bibnamefont
  {Kirschner}},\ }\href {\doibase 10.1038/312237a0} {\bibfield  {journal}
  {\bibinfo  {journal} {Nature}\ }\textbf {\bibinfo {volume} {312}},\ \bibinfo
  {pages} {237} (\bibinfo {year} {1984})}\BibitemShut {NoStop}%
\bibitem [{\citenamefont {Akhmanova}\ and\ \citenamefont
  {Steinmetz}(2015)}]{Akhmanova2015}%
  \BibitemOpen
  \bibfield  {author} {\bibinfo {author} {\bibfnamefont {A.}~\bibnamefont
  {Akhmanova}}\ and\ \bibinfo {author} {\bibfnamefont {M.~O.}\ \bibnamefont
  {Steinmetz}},\ }\href {\doibase 10.1038/nrm4084} {\bibfield  {journal}
  {\bibinfo  {journal} {Nat. Rev. Mol. Cell Biol.}\ }\textbf {\bibinfo {volume}
  {16}},\ \bibinfo {pages} {711} (\bibinfo {year} {2015})}\BibitemShut
  {NoStop}%
\bibitem [{\citenamefont {Cleary}\ and\ \citenamefont
  {Hancock}(2021)}]{Cleary2021}%
  \BibitemOpen
  \bibfield  {author} {\bibinfo {author} {\bibfnamefont {J.~M.}\ \bibnamefont
  {Cleary}}\ and\ \bibinfo {author} {\bibfnamefont {W.~O.}\ \bibnamefont
  {Hancock}},\ }\href {\doibase 10.1016/j.cub.2021.02.035} {\bibfield
  {journal} {\bibinfo  {journal} {Curr. Biol.}\ }\textbf {\bibinfo {volume}
  {31}},\ \bibinfo {pages} {R560} (\bibinfo {year} {2021})}\BibitemShut
  {NoStop}%
\bibitem [{\citenamefont {Dye}\ \emph {et~al.}(1992)\citenamefont {Dye},
  \citenamefont {Flicker}, \citenamefont {Lien},\ and\ \citenamefont
  {Williams}}]{Dye1992}%
  \BibitemOpen
  \bibfield  {author} {\bibinfo {author} {\bibfnamefont {R.~B.}\ \bibnamefont
  {Dye}}, \bibinfo {author} {\bibfnamefont {P.~F.}\ \bibnamefont {Flicker}},
  \bibinfo {author} {\bibfnamefont {D.~Y.}\ \bibnamefont {Lien}}, \ and\
  \bibinfo {author} {\bibfnamefont {R.~C.}\ \bibnamefont {Williams}},\ }\href
  {\doibase 10.1002/cm.970210302} {\bibfield  {journal} {\bibinfo  {journal}
  {Cell Mot. Cytoskel.}\ }\textbf {\bibinfo {volume} {21}},\ \bibinfo {pages}
  {171} (\bibinfo {year} {1992})}\BibitemShut {NoStop}%
\bibitem [{\citenamefont {Schaedel}\ \emph {et~al.}(2015)\citenamefont
  {Schaedel}, \citenamefont {John}, \citenamefont {Gaillard}, \citenamefont
  {Nachury}, \citenamefont {Blanchoin},\ and\ \citenamefont
  {Th{\'e}ry}}]{Schaedel2015}%
  \BibitemOpen
  \bibfield  {author} {\bibinfo {author} {\bibfnamefont {L.}~\bibnamefont
  {Schaedel}}, \bibinfo {author} {\bibfnamefont {K.}~\bibnamefont {John}},
  \bibinfo {author} {\bibfnamefont {J.}~\bibnamefont {Gaillard}}, \bibinfo
  {author} {\bibfnamefont {M.~V.}\ \bibnamefont {Nachury}}, \bibinfo {author}
  {\bibfnamefont {L.}~\bibnamefont {Blanchoin}}, \ and\ \bibinfo {author}
  {\bibfnamefont {M.}~\bibnamefont {Th{\'e}ry}},\ }\href {\doibase
  10.1038/nmat4396} {\bibfield  {journal} {\bibinfo  {journal} {Nat. Mater.}\
  }\textbf {\bibinfo {volume} {14}},\ \bibinfo {pages} {1156} (\bibinfo {year}
  {2015})}\BibitemShut {NoStop}%
\bibitem [{\citenamefont {Aumeier}\ \emph {et~al.}(2016)\citenamefont
  {Aumeier}, \citenamefont {Schaedel}, \citenamefont {Gaillard}, \citenamefont
  {John}, \citenamefont {Blanchoin},\ and\ \citenamefont
  {Th{\'e}ry}}]{Aumeier2016}%
  \BibitemOpen
  \bibfield  {author} {\bibinfo {author} {\bibfnamefont {C.}~\bibnamefont
  {Aumeier}}, \bibinfo {author} {\bibfnamefont {L.}~\bibnamefont {Schaedel}},
  \bibinfo {author} {\bibfnamefont {J.}~\bibnamefont {Gaillard}}, \bibinfo
  {author} {\bibfnamefont {K.}~\bibnamefont {John}}, \bibinfo {author}
  {\bibfnamefont {L.}~\bibnamefont {Blanchoin}}, \ and\ \bibinfo {author}
  {\bibfnamefont {M.}~\bibnamefont {Th{\'e}ry}},\ }\href {\doibase
  10.1038/ncb3406} {\bibfield  {journal} {\bibinfo  {journal} {Nat. Cell
  Biol.}\ }\textbf {\bibinfo {volume} {18}},\ \bibinfo {pages} {1054} (\bibinfo
  {year} {2016})}\BibitemShut {NoStop}%
\bibitem [{\citenamefont {Vemu}\ \emph {et~al.}(2018)\citenamefont {Vemu},
  \citenamefont {Szczesna}, \citenamefont {Zehr}, \citenamefont {Spector},
  \citenamefont {Grigorieff}, \citenamefont {Deaconescu},\ and\ \citenamefont
  {Roll-Mecak}}]{Vemu2018}%
  \BibitemOpen
  \bibfield  {author} {\bibinfo {author} {\bibfnamefont {A.}~\bibnamefont
  {Vemu}}, \bibinfo {author} {\bibfnamefont {E.}~\bibnamefont {Szczesna}},
  \bibinfo {author} {\bibfnamefont {E.~A.}\ \bibnamefont {Zehr}}, \bibinfo
  {author} {\bibfnamefont {J.~O.}\ \bibnamefont {Spector}}, \bibinfo {author}
  {\bibfnamefont {N.}~\bibnamefont {Grigorieff}}, \bibinfo {author}
  {\bibfnamefont {A.~M.}\ \bibnamefont {Deaconescu}}, \ and\ \bibinfo {author}
  {\bibfnamefont {A.}~\bibnamefont {Roll-Mecak}},\ }\href {\doibase
  10.1126/science.aau1504} {\bibfield  {journal} {\bibinfo  {journal}
  {Science}\ }\textbf {\bibinfo {volume} {361}},\ \bibinfo {pages} {eaau1504}
  (\bibinfo {year} {2018})}\BibitemShut {NoStop}%
\bibitem [{\citenamefont {Schaedel}\ \emph {et~al.}(2019)\citenamefont
  {Schaedel}, \citenamefont {Triclin}, \citenamefont {Chr{\'e}tien},
  \citenamefont {Abrieu}, \citenamefont {Aumeier}, \citenamefont {Gaillard},
  \citenamefont {Blanchoin}, \citenamefont {Th{\'e}ry},\ and\ \citenamefont
  {John}}]{Schaedel2019}%
  \BibitemOpen
  \bibfield  {author} {\bibinfo {author} {\bibfnamefont {L.}~\bibnamefont
  {Schaedel}}, \bibinfo {author} {\bibfnamefont {S.}~\bibnamefont {Triclin}},
  \bibinfo {author} {\bibfnamefont {D.}~\bibnamefont {Chr{\'e}tien}}, \bibinfo
  {author} {\bibfnamefont {A.}~\bibnamefont {Abrieu}}, \bibinfo {author}
  {\bibfnamefont {C.}~\bibnamefont {Aumeier}}, \bibinfo {author} {\bibfnamefont
  {J.}~\bibnamefont {Gaillard}}, \bibinfo {author} {\bibfnamefont
  {L.}~\bibnamefont {Blanchoin}}, \bibinfo {author} {\bibfnamefont
  {M.}~\bibnamefont {Th{\'e}ry}}, \ and\ \bibinfo {author} {\bibfnamefont
  {K.}~\bibnamefont {John}},\ }\href {\doibase 10.1038/s41567-019-0542-4}
  {\bibfield  {journal} {\bibinfo  {journal} {Nat. Phys.}\ }\textbf {\bibinfo
  {volume} {15}},\ \bibinfo {pages} {830} (\bibinfo {year} {2019})}\BibitemShut
  {NoStop}%
\bibitem [{\citenamefont {Triclin}\ \emph {et~al.}(2021)\citenamefont
  {Triclin}, \citenamefont {Inoue}, \citenamefont {Gaillard}, \citenamefont
  {Htet}, \citenamefont {DeSantis}, \citenamefont {Portran}, \citenamefont
  {Derivery}, \citenamefont {Aumeier}, \citenamefont {Schaedel}, \citenamefont
  {John} \emph {et~al.}}]{Triclin2021}%
  \BibitemOpen
  \bibfield  {author} {\bibinfo {author} {\bibfnamefont {S.}~\bibnamefont
  {Triclin}}, \bibinfo {author} {\bibfnamefont {D.}~\bibnamefont {Inoue}},
  \bibinfo {author} {\bibfnamefont {J.}~\bibnamefont {Gaillard}}, \bibinfo
  {author} {\bibfnamefont {Z.~M.}\ \bibnamefont {Htet}}, \bibinfo {author}
  {\bibfnamefont {M.~E.}\ \bibnamefont {DeSantis}}, \bibinfo {author}
  {\bibfnamefont {D.}~\bibnamefont {Portran}}, \bibinfo {author} {\bibfnamefont
  {E.}~\bibnamefont {Derivery}}, \bibinfo {author} {\bibfnamefont
  {C.}~\bibnamefont {Aumeier}}, \bibinfo {author} {\bibfnamefont
  {L.}~\bibnamefont {Schaedel}}, \bibinfo {author} {\bibfnamefont
  {K.}~\bibnamefont {John}},  \emph {et~al.},\ }\href {\doibase
  10.1038/s41563-020-00905-0} {\bibfield  {journal} {\bibinfo  {journal} {Nat.
  Mater.}\ }\textbf {\bibinfo {volume} {20}},\ \bibinfo {pages} {883} (\bibinfo
  {year} {2021})}\BibitemShut {NoStop}%
\bibitem [{\citenamefont {Alexandrova}\ \emph {et~al.}(2022)\citenamefont
  {Alexandrova}, \citenamefont {Anisimov}, \citenamefont {Zaitsev},
  \citenamefont {Mustyatsa}, \citenamefont {Popov}, \citenamefont
  {Ataullakhanov},\ and\ \citenamefont {Gudimchuk}}]{Alexandrova2022}%
  \BibitemOpen
  \bibfield  {author} {\bibinfo {author} {\bibfnamefont {V.~V.}\ \bibnamefont
  {Alexandrova}}, \bibinfo {author} {\bibfnamefont {M.~N.}\ \bibnamefont
  {Anisimov}}, \bibinfo {author} {\bibfnamefont {A.~V.}\ \bibnamefont
  {Zaitsev}}, \bibinfo {author} {\bibfnamefont {V.~V.}\ \bibnamefont
  {Mustyatsa}}, \bibinfo {author} {\bibfnamefont {V.~V.}\ \bibnamefont
  {Popov}}, \bibinfo {author} {\bibfnamefont {F.~I.}\ \bibnamefont
  {Ataullakhanov}}, \ and\ \bibinfo {author} {\bibfnamefont {N.~B.}\
  \bibnamefont {Gudimchuk}},\ }\href {\doibase 10.1073/pnas.2208294119}
  {\bibfield  {journal} {\bibinfo  {journal} {Proc. Natl. Acad. Sci. USA}\
  }\textbf {\bibinfo {volume} {119}} (\bibinfo {year} {2022}),\
  10.1073/pnas.2208294119}\BibitemShut {NoStop}%
\bibitem [{\citenamefont {Andreu-Carbó}\ \emph {et~al.}(2022)\citenamefont
  {Andreu-Carbó}, \citenamefont {Fernandes}, \citenamefont {Velluz},
  \citenamefont {Kruse},\ and\ \citenamefont {Aumeier}}]{AndreuCarbo2022}%
  \BibitemOpen
  \bibfield  {author} {\bibinfo {author} {\bibfnamefont {M.}~\bibnamefont
  {Andreu-Carbó}}, \bibinfo {author} {\bibfnamefont {S.}~\bibnamefont
  {Fernandes}}, \bibinfo {author} {\bibfnamefont {M.-C.}\ \bibnamefont
  {Velluz}}, \bibinfo {author} {\bibfnamefont {K.}~\bibnamefont {Kruse}}, \
  and\ \bibinfo {author} {\bibfnamefont {C.}~\bibnamefont {Aumeier}},\ }\href
  {\doibase https://doi.org/10.1016/j.devcel.2021.11.019} {\bibfield  {journal}
  {\bibinfo  {journal} {Dev. Cell}\ }\textbf {\bibinfo {volume} {57}},\
  \bibinfo {pages} {5} (\bibinfo {year} {2022})}\BibitemShut {NoStop}%
\bibitem [{\citenamefont {Budaitis}\ \emph {et~al.}(2022)\citenamefont
  {Budaitis}, \citenamefont {Badieyan}, \citenamefont {Yue}, \citenamefont
  {Blasius}, \citenamefont {Reinemann}, \citenamefont {Lang}, \citenamefont
  {Cianfrocco},\ and\ \citenamefont {Verhey}}]{Budaitis2022}%
  \BibitemOpen
  \bibfield  {author} {\bibinfo {author} {\bibfnamefont {B.~G.}\ \bibnamefont
  {Budaitis}}, \bibinfo {author} {\bibfnamefont {S.}~\bibnamefont {Badieyan}},
  \bibinfo {author} {\bibfnamefont {Y.}~\bibnamefont {Yue}}, \bibinfo {author}
  {\bibfnamefont {T.~L.}\ \bibnamefont {Blasius}}, \bibinfo {author}
  {\bibfnamefont {D.~N.}\ \bibnamefont {Reinemann}}, \bibinfo {author}
  {\bibfnamefont {M.~J.}\ \bibnamefont {Lang}}, \bibinfo {author}
  {\bibfnamefont {M.~A.}\ \bibnamefont {Cianfrocco}}, \ and\ \bibinfo {author}
  {\bibfnamefont {K.~J.}\ \bibnamefont {Verhey}},\ }\href {\doibase
  10.1016/j.cub.2022.04.020} {\bibfield  {journal} {\bibinfo  {journal} {Curr.
  Biol.}\ }\textbf {\bibinfo {volume} {32}},\ \bibinfo {pages} {2416} (\bibinfo
  {year} {2022})}\BibitemShut {NoStop}%
\bibitem [{\citenamefont {Gazzola}\ \emph {et~al.}(2023)\citenamefont
  {Gazzola}, \citenamefont {Schaeffer}, \citenamefont {Butler-Hallissey},
  \citenamefont {Friedl}, \citenamefont {Vianay}, \citenamefont {Gaillard},
  \citenamefont {Leterrier}, \citenamefont {Blanchoin},\ and\ \citenamefont
  {Théry}}]{Gazzola2023}%
  \BibitemOpen
  \bibfield  {author} {\bibinfo {author} {\bibfnamefont {M.}~\bibnamefont
  {Gazzola}}, \bibinfo {author} {\bibfnamefont {A.}~\bibnamefont {Schaeffer}},
  \bibinfo {author} {\bibfnamefont {C.}~\bibnamefont {Butler-Hallissey}},
  \bibinfo {author} {\bibfnamefont {K.}~\bibnamefont {Friedl}}, \bibinfo
  {author} {\bibfnamefont {B.}~\bibnamefont {Vianay}}, \bibinfo {author}
  {\bibfnamefont {J.}~\bibnamefont {Gaillard}}, \bibinfo {author}
  {\bibfnamefont {C.}~\bibnamefont {Leterrier}}, \bibinfo {author}
  {\bibfnamefont {L.}~\bibnamefont {Blanchoin}}, \ and\ \bibinfo {author}
  {\bibfnamefont {M.}~\bibnamefont {Théry}},\ }\href {\doibase
  10.1016/j.cub.2022.11.060} {\bibfield  {journal} {\bibinfo  {journal} {Curr.
  Biol.}\ }\textbf {\bibinfo {volume} {33}},\ \bibinfo {pages} {122} (\bibinfo
  {year} {2023})}\BibitemShut {NoStop}%
\bibitem [{\citenamefont {Biswas}\ \emph {et~al.}(2025)\citenamefont {Biswas},
  \citenamefont {Grover}, \citenamefont {Reuther}, \citenamefont {Poojari},
  \citenamefont {Shaebani}, \citenamefont {Nandakumar}, \citenamefont
  {Grünewald}, \citenamefont {Zablotsky}, \citenamefont {Hub}, \citenamefont
  {Diez}, \citenamefont {John},\ and\ \citenamefont {Schaedel}}]{Biswas2025}%
  \BibitemOpen
  \bibfield  {author} {\bibinfo {author} {\bibfnamefont {S.}~\bibnamefont
  {Biswas}}, \bibinfo {author} {\bibfnamefont {R.}~\bibnamefont {Grover}},
  \bibinfo {author} {\bibfnamefont {C.}~\bibnamefont {Reuther}}, \bibinfo
  {author} {\bibfnamefont {C.~S.}\ \bibnamefont {Poojari}}, \bibinfo {author}
  {\bibfnamefont {R.}~\bibnamefont {Shaebani}}, \bibinfo {author}
  {\bibfnamefont {S.}~\bibnamefont {Nandakumar}}, \bibinfo {author}
  {\bibfnamefont {M.}~\bibnamefont {Grünewald}}, \bibinfo {author}
  {\bibfnamefont {A.}~\bibnamefont {Zablotsky}}, \bibinfo {author}
  {\bibfnamefont {J.~S.}\ \bibnamefont {Hub}}, \bibinfo {author} {\bibfnamefont
  {S.}~\bibnamefont {Diez}}, \bibinfo {author} {\bibfnamefont {K.}~\bibnamefont
  {John}}, \ and\ \bibinfo {author} {\bibfnamefont {L.}~\bibnamefont
  {Schaedel}},\ }\href {\doibase 10.1038/s41567-025-03003-7} {\bibfield
  {journal} {\bibinfo  {journal} {Nature Physics}\ }\textbf {\bibinfo {volume}
  {21}},\ \bibinfo {pages} {1616} (\bibinfo {year} {2025})}\BibitemShut
  {NoStop}%
\bibitem [{\citenamefont {McNally}\ and\ \citenamefont
  {Roll-Mecak}(2018)}]{McNally2018}%
  \BibitemOpen
  \bibfield  {author} {\bibinfo {author} {\bibfnamefont {F.~J.}\ \bibnamefont
  {McNally}}\ and\ \bibinfo {author} {\bibfnamefont {A.}~\bibnamefont
  {Roll-Mecak}},\ }\href {\doibase 10.1083/jcb.201612104} {\bibfield  {journal}
  {\bibinfo  {journal} {J. Cell Biol.}\ }\textbf {\bibinfo {volume} {217}},\
  \bibinfo {pages} {4057} (\bibinfo {year} {2018})}\BibitemShut {NoStop}%
\bibitem [{\citenamefont {Tang-Schomer}\ \emph {et~al.}(2009)\citenamefont
  {Tang-Schomer}, \citenamefont {Patel}, \citenamefont {Baas},\ and\
  \citenamefont {Smith}}]{TangSchomer2009}%
  \BibitemOpen
  \bibfield  {author} {\bibinfo {author} {\bibfnamefont {M.~D.}\ \bibnamefont
  {Tang-Schomer}}, \bibinfo {author} {\bibfnamefont {A.~R.}\ \bibnamefont
  {Patel}}, \bibinfo {author} {\bibfnamefont {P.~W.}\ \bibnamefont {Baas}}, \
  and\ \bibinfo {author} {\bibfnamefont {D.~H.}\ \bibnamefont {Smith}},\ }\href
  {\doibase 10.1096/fj.09-142844} {\bibfield  {journal} {\bibinfo  {journal}
  {FASEB J.}\ }\textbf {\bibinfo {volume} {24}},\ \bibinfo {pages} {1401}
  (\bibinfo {year} {2009})}\BibitemShut {NoStop}%
\bibitem [{\citenamefont {Nandakumar}\ \emph {et~al.}(2025)\citenamefont
  {Nandakumar}, \citenamefont {Bosche}, \citenamefont {Wieczorek},
  \citenamefont {Albrecht}, \citenamefont {König}, \citenamefont {Grünewald},
  \citenamefont {Santen}, \citenamefont {Diez}, \citenamefont {Shaebani},\ and\
  \citenamefont {Schaedel}}]{Nandakumar2025}%
  \BibitemOpen
  \bibfield  {author} {\bibinfo {author} {\bibfnamefont {S.}~\bibnamefont
  {Nandakumar}}, \bibinfo {author} {\bibfnamefont {J.}~\bibnamefont {Bosche}},
  \bibinfo {author} {\bibfnamefont {M.}~\bibnamefont {Wieczorek}}, \bibinfo
  {author} {\bibfnamefont {C.~M.}\ \bibnamefont {Albrecht}}, \bibinfo {author}
  {\bibfnamefont {B.}~\bibnamefont {König}}, \bibinfo {author} {\bibfnamefont
  {M.}~\bibnamefont {Grünewald}}, \bibinfo {author} {\bibfnamefont
  {L.}~\bibnamefont {Santen}}, \bibinfo {author} {\bibfnamefont
  {S.}~\bibnamefont {Diez}}, \bibinfo {author} {\bibfnamefont {R.}~\bibnamefont
  {Shaebani}}, \ and\ \bibinfo {author} {\bibfnamefont {L.}~\bibnamefont
  {Schaedel}},\ }\href {\doibase 10.1101/2025.09.08.672697} {\bibfield
  {journal} {\bibinfo  {journal} {BioRxiv}\ } (\bibinfo {year} {2025}),\
  10.1101/2025.09.08.672697}\BibitemShut {NoStop}%
\bibitem [{\citenamefont {Guyomar}\ \emph {et~al.}(2022)\citenamefont
  {Guyomar}, \citenamefont {Bousquet}, \citenamefont {Ku}, \citenamefont
  {Heumann}, \citenamefont {Guilloux}, \citenamefont {Gaillard}, \citenamefont
  {Heichette}, \citenamefont {Duchesne}, \citenamefont {Steinmetz},
  \citenamefont {Gibeaux},\ and\ \citenamefont {Chrétien}}]{Guyomar2022}%
  \BibitemOpen
  \bibfield  {author} {\bibinfo {author} {\bibfnamefont {C.}~\bibnamefont
  {Guyomar}}, \bibinfo {author} {\bibfnamefont {C.}~\bibnamefont {Bousquet}},
  \bibinfo {author} {\bibfnamefont {S.}~\bibnamefont {Ku}}, \bibinfo {author}
  {\bibfnamefont {J.~M.}\ \bibnamefont {Heumann}}, \bibinfo {author}
  {\bibfnamefont {G.}~\bibnamefont {Guilloux}}, \bibinfo {author}
  {\bibfnamefont {N.}~\bibnamefont {Gaillard}}, \bibinfo {author}
  {\bibfnamefont {C.}~\bibnamefont {Heichette}}, \bibinfo {author}
  {\bibfnamefont {L.}~\bibnamefont {Duchesne}}, \bibinfo {author}
  {\bibfnamefont {M.~O.}\ \bibnamefont {Steinmetz}}, \bibinfo {author}
  {\bibfnamefont {R.}~\bibnamefont {Gibeaux}}, \ and\ \bibinfo {author}
  {\bibfnamefont {D.}~\bibnamefont {Chrétien}},\ }\href {\doibase
  10.7554/elife.83021} {\bibfield  {journal} {\bibinfo  {journal} {eLife}\
  }\textbf {\bibinfo {volume} {11}},\ \bibinfo {pages} {e83021} (\bibinfo
  {year} {2022})}\BibitemShut {NoStop}%
\bibitem [{\citenamefont {Lukkien}\ \emph {et~al.}(1998)\citenamefont
  {Lukkien}, \citenamefont {Segers}, \citenamefont {Hilbers}, \citenamefont
  {Gelten},\ and\ \citenamefont {Jansen}}]{Lukkien1998}%
  \BibitemOpen
  \bibfield  {author} {\bibinfo {author} {\bibfnamefont {J.}~\bibnamefont
  {Lukkien}}, \bibinfo {author} {\bibfnamefont {J.}~\bibnamefont {Segers}},
  \bibinfo {author} {\bibfnamefont {P.}~\bibnamefont {Hilbers}}, \bibinfo
  {author} {\bibfnamefont {R.}~\bibnamefont {Gelten}}, \ and\ \bibinfo {author}
  {\bibfnamefont {A.}~\bibnamefont {Jansen}},\ }\href {\doibase
  doi:10.1103/PhysRevE.58.2598} {\bibfield  {journal} {\bibinfo  {journal}
  {Phys. Rev. E}\ }\textbf {\bibinfo {volume} {58}},\ \bibinfo {pages} {2598}
  (\bibinfo {year} {1998})}\BibitemShut {NoStop}%
\bibitem [{Sup()}]{SupMat}%
  \BibitemOpen
  \href@noop {} {}\bibinfo {note} {See Supplemental Material [url] for details
  of the kinetic Monte Carlo simulations, the calibration of the time scale
  $\tau$, additional simulations of the damge size and time to fracture, and a
  discussion of literature values for $A$ and $\Delta G_\text{b}$.}\BibitemShut
  {Stop}%
\bibitem [{\citenamefont {VanBuren}\ \emph {et~al.}(2002)\citenamefont
  {VanBuren}, \citenamefont {Odde},\ and\ \citenamefont
  {Cassimeris}}]{vanBuren2002}%
  \BibitemOpen
  \bibfield  {author} {\bibinfo {author} {\bibfnamefont {V.}~\bibnamefont
  {VanBuren}}, \bibinfo {author} {\bibfnamefont {D.~J.}\ \bibnamefont {Odde}},
  \ and\ \bibinfo {author} {\bibfnamefont {L.}~\bibnamefont {Cassimeris}},\
  }\href {\doibase 10.1073/pnas.092504999} {\bibfield  {journal} {\bibinfo
  {journal} {Proc. Natl. Acad. Sci. USA}\ }\textbf {\bibinfo {volume} {99}},\
  \bibinfo {pages} {6035} (\bibinfo {year} {2002})}\BibitemShut {NoStop}%
\bibitem [{\citenamefont {Gardner}\ \emph {et~al.}(2011)\citenamefont
  {Gardner}, \citenamefont {Charlebois}, \citenamefont {Jánosi}, \citenamefont
  {Howard}, \citenamefont {Hunt},\ and\ \citenamefont {Odde}}]{Gardner2011}%
  \BibitemOpen
  \bibfield  {author} {\bibinfo {author} {\bibfnamefont {M.}~\bibnamefont
  {Gardner}}, \bibinfo {author} {\bibfnamefont {B.}~\bibnamefont {Charlebois}},
  \bibinfo {author} {\bibfnamefont {I.}~\bibnamefont {Jánosi}}, \bibinfo
  {author} {\bibfnamefont {J.}~\bibnamefont {Howard}}, \bibinfo {author}
  {\bibfnamefont {A.}~\bibnamefont {Hunt}}, \ and\ \bibinfo {author}
  {\bibfnamefont {D.}~\bibnamefont {Odde}},\ }\href {\doibase
  10.1016/j.cell.2011.06.053} {\bibfield  {journal} {\bibinfo  {journal}
  {Cell}\ }\textbf {\bibinfo {volume} {146}},\ \bibinfo {pages} {582} (\bibinfo
  {year} {2011})}\BibitemShut {NoStop}%
\bibitem [{\citenamefont {Margolin}\ \emph {et~al.}(2012)\citenamefont
  {Margolin}, \citenamefont {Gregoretti}, \citenamefont {Cickovski},
  \citenamefont {Li}, \citenamefont {Shi}, \citenamefont {Alber},\ and\
  \citenamefont {Goodson}}]{Margolin2012}%
  \BibitemOpen
  \bibfield  {author} {\bibinfo {author} {\bibfnamefont {G.}~\bibnamefont
  {Margolin}}, \bibinfo {author} {\bibfnamefont {I.~V.}\ \bibnamefont
  {Gregoretti}}, \bibinfo {author} {\bibfnamefont {T.~M.}\ \bibnamefont
  {Cickovski}}, \bibinfo {author} {\bibfnamefont {C.}~\bibnamefont {Li}},
  \bibinfo {author} {\bibfnamefont {W.}~\bibnamefont {Shi}}, \bibinfo {author}
  {\bibfnamefont {M.~S.}\ \bibnamefont {Alber}}, \ and\ \bibinfo {author}
  {\bibfnamefont {H.~V.}\ \bibnamefont {Goodson}},\ }\href {\doibase
  10.1091/mbc.e11-08-0688} {\bibfield  {journal} {\bibinfo  {journal} {Mol.
  Biol. Cell}\ }\textbf {\bibinfo {volume} {23}},\ \bibinfo {pages} {642}
  (\bibinfo {year} {2012})}\BibitemShut {NoStop}%
\bibitem [{\citenamefont {Piedra}\ \emph {et~al.}(2016)\citenamefont {Piedra},
  \citenamefont {Kim}, \citenamefont {Garza}, \citenamefont {Geyer},
  \citenamefont {Burns}, \citenamefont {Ye},\ and\ \citenamefont
  {Rice}}]{Piedra2016}%
  \BibitemOpen
  \bibfield  {author} {\bibinfo {author} {\bibfnamefont {F.-A.}\ \bibnamefont
  {Piedra}}, \bibinfo {author} {\bibfnamefont {T.}~\bibnamefont {Kim}},
  \bibinfo {author} {\bibfnamefont {E.~S.}\ \bibnamefont {Garza}}, \bibinfo
  {author} {\bibfnamefont {E.~A.}\ \bibnamefont {Geyer}}, \bibinfo {author}
  {\bibfnamefont {A.}~\bibnamefont {Burns}}, \bibinfo {author} {\bibfnamefont
  {X.}~\bibnamefont {Ye}}, \ and\ \bibinfo {author} {\bibfnamefont {L.~M.}\
  \bibnamefont {Rice}},\ }\href@noop {} {\bibfield  {journal} {\bibinfo
  {journal} {Mol. Biol. Cell}\ }\textbf {\bibinfo {volume} {27}},\ \bibinfo
  {pages} {3515} (\bibinfo {year} {2016})}\BibitemShut {NoStop}%
\bibitem [{\citenamefont {Chaaban}\ \emph {et~al.}(2018)\citenamefont
  {Chaaban}, \citenamefont {Jariwala}, \citenamefont {Hsu}, \citenamefont
  {Redemann}, \citenamefont {Kollman}, \citenamefont {Müller-Reichert},
  \citenamefont {Sept}, \citenamefont {Bui},\ and\ \citenamefont
  {Brouhard}}]{Chaaban2018}%
  \BibitemOpen
  \bibfield  {author} {\bibinfo {author} {\bibfnamefont {S.}~\bibnamefont
  {Chaaban}}, \bibinfo {author} {\bibfnamefont {S.}~\bibnamefont {Jariwala}},
  \bibinfo {author} {\bibfnamefont {C.-T.}\ \bibnamefont {Hsu}}, \bibinfo
  {author} {\bibfnamefont {S.}~\bibnamefont {Redemann}}, \bibinfo {author}
  {\bibfnamefont {J.~M.}\ \bibnamefont {Kollman}}, \bibinfo {author}
  {\bibfnamefont {T.}~\bibnamefont {Müller-Reichert}}, \bibinfo {author}
  {\bibfnamefont {D.}~\bibnamefont {Sept}}, \bibinfo {author} {\bibfnamefont
  {K.~H.}\ \bibnamefont {Bui}}, \ and\ \bibinfo {author} {\bibfnamefont
  {G.~J.}\ \bibnamefont {Brouhard}},\ }\href {\doibase
  10.1016/j.devcel.2018.08.023} {\bibfield  {journal} {\bibinfo  {journal}
  {Dev. Cell}\ }\textbf {\bibinfo {volume} {47}},\ \bibinfo {pages} {191}
  (\bibinfo {year} {2018})}\BibitemShut {NoStop}%
\bibitem [{\citenamefont {Mickolajczyk}\ \emph {et~al.}(2019)\citenamefont
  {Mickolajczyk}, \citenamefont {Geyer}, \citenamefont {Kim}, \citenamefont
  {Rice},\ and\ \citenamefont {Hancock}}]{Mickolajczyk2019}%
  \BibitemOpen
  \bibfield  {author} {\bibinfo {author} {\bibfnamefont {K.~J.}\ \bibnamefont
  {Mickolajczyk}}, \bibinfo {author} {\bibfnamefont {E.~A.}\ \bibnamefont
  {Geyer}}, \bibinfo {author} {\bibfnamefont {T.}~\bibnamefont {Kim}}, \bibinfo
  {author} {\bibfnamefont {L.~M.}\ \bibnamefont {Rice}}, \ and\ \bibinfo
  {author} {\bibfnamefont {W.~O.}\ \bibnamefont {Hancock}},\ }\href {\doibase
  10.1073/pnas.1815823116} {\bibfield  {journal} {\bibinfo  {journal} {Proc.
  Natl. Acad. Sci. USA}\ }\textbf {\bibinfo {volume} {116}},\ \bibinfo {pages}
  {7314} (\bibinfo {year} {2019})}\BibitemShut {NoStop}%
\bibitem [{\citenamefont {Thawani}\ \emph {et~al.}(2020)\citenamefont
  {Thawani}, \citenamefont {Rale}, \citenamefont {Coudray}, \citenamefont
  {Bhabha}, \citenamefont {Stone}, \citenamefont {Shaevitz},\ and\
  \citenamefont {Petry}}]{Thawani2020}%
  \BibitemOpen
  \bibfield  {author} {\bibinfo {author} {\bibfnamefont {A.}~\bibnamefont
  {Thawani}}, \bibinfo {author} {\bibfnamefont {M.~J.}\ \bibnamefont {Rale}},
  \bibinfo {author} {\bibfnamefont {N.}~\bibnamefont {Coudray}}, \bibinfo
  {author} {\bibfnamefont {G.}~\bibnamefont {Bhabha}}, \bibinfo {author}
  {\bibfnamefont {H.~A.}\ \bibnamefont {Stone}}, \bibinfo {author}
  {\bibfnamefont {J.~W.}\ \bibnamefont {Shaevitz}}, \ and\ \bibinfo {author}
  {\bibfnamefont {S.}~\bibnamefont {Petry}},\ }\href {\doibase
  10.7554/elife.54253} {\bibfield  {journal} {\bibinfo  {journal} {eLife}\
  }\textbf {\bibinfo {volume} {9}} (\bibinfo {year} {2020}),\
  10.7554/elife.54253}\BibitemShut {NoStop}%
\bibitem [{\citenamefont {McCormick}\ \emph {et~al.}(2024)\citenamefont
  {McCormick}, \citenamefont {Cleary}, \citenamefont {Hancock},\ and\
  \citenamefont {Rice}}]{McCormick2024}%
  \BibitemOpen
  \bibfield  {author} {\bibinfo {author} {\bibfnamefont {L.~A.}\ \bibnamefont
  {McCormick}}, \bibinfo {author} {\bibfnamefont {J.~M.}\ \bibnamefont
  {Cleary}}, \bibinfo {author} {\bibfnamefont {W.~O.}\ \bibnamefont {Hancock}},
  \ and\ \bibinfo {author} {\bibfnamefont {L.~M.}\ \bibnamefont {Rice}},\
  }\href {\doibase 10.7554/elife.89231} {\bibfield  {journal} {\bibinfo
  {journal} {eLife}\ }\textbf {\bibinfo {volume} {12}} (\bibinfo {year}
  {2024}),\ 10.7554/elife.89231}\BibitemShut {NoStop}%
\bibitem [{\citenamefont {Kondepudi}\ and\ \citenamefont
  {Prigogine}(2014)}]{Kondepudi2014}%
  \BibitemOpen
  \bibfield  {author} {\bibinfo {author} {\bibfnamefont {D.}~\bibnamefont
  {Kondepudi}}\ and\ \bibinfo {author} {\bibfnamefont {I.}~\bibnamefont
  {Prigogine}},\ }\href {\doibase 10.1002/9781118698723} {\emph {\bibinfo
  {title} {Modern Thermodynamics: From Heat Engines to Dissipative
  Structures}}}\ (\bibinfo  {publisher} {Wiley},\ \bibinfo {year}
  {2014})\BibitemShut {NoStop}%
\bibitem [{\citenamefont {Cassimeris}\ \emph {et~al.}(1986)\citenamefont
  {Cassimeris}, \citenamefont {Wadsworth},\ and\ \citenamefont
  {Salmon}}]{Cassimeris1986}%
  \BibitemOpen
  \bibfield  {author} {\bibinfo {author} {\bibfnamefont {L.~U.}\ \bibnamefont
  {Cassimeris}}, \bibinfo {author} {\bibfnamefont {P.}~\bibnamefont
  {Wadsworth}}, \ and\ \bibinfo {author} {\bibfnamefont {E.~D.}\ \bibnamefont
  {Salmon}},\ }\href {\doibase 10.1083/jcb.102.6.2023} {\bibfield  {journal}
  {\bibinfo  {journal} {J. Cell Biol.}\ }\textbf {\bibinfo {volume} {102}},\
  \bibinfo {pages} {2023} (\bibinfo {year} {1986})}\BibitemShut {NoStop}%
\bibitem [{\citenamefont {Walker}\ \emph {et~al.}(1988)\citenamefont {Walker},
  \citenamefont {O'Brien}, \citenamefont {Pryer}, \citenamefont {Soboeiro},
  \citenamefont {Voter}, \citenamefont {Erickson},\ and\ \citenamefont
  {Salmon}}]{Walker1988}%
  \BibitemOpen
  \bibfield  {author} {\bibinfo {author} {\bibfnamefont {R.~A.}\ \bibnamefont
  {Walker}}, \bibinfo {author} {\bibfnamefont {E.~T.}\ \bibnamefont {O'Brien}},
  \bibinfo {author} {\bibfnamefont {N.~K.}\ \bibnamefont {Pryer}}, \bibinfo
  {author} {\bibfnamefont {M.~F.}\ \bibnamefont {Soboeiro}}, \bibinfo {author}
  {\bibfnamefont {W.~A.}\ \bibnamefont {Voter}}, \bibinfo {author}
  {\bibfnamefont {H.~P.}\ \bibnamefont {Erickson}}, \ and\ \bibinfo {author}
  {\bibfnamefont {E.~D.}\ \bibnamefont {Salmon}},\ }\href {\doibase
  10.1083/jcb.107.4.1437} {\bibfield  {journal} {\bibinfo  {journal} {J. Cell
  Biol.}\ }\textbf {\bibinfo {volume} {107}},\ \bibinfo {pages} {1437}
  (\bibinfo {year} {1988})}\BibitemShut {NoStop}%
\bibitem [{\citenamefont {Verde}\ \emph {et~al.}(1992)\citenamefont {Verde},
  \citenamefont {Dogterom}, \citenamefont {Stelzer}, \citenamefont {Karsenti},\
  and\ \citenamefont {Leibler}}]{Verde1992}%
  \BibitemOpen
  \bibfield  {author} {\bibinfo {author} {\bibfnamefont {F.}~\bibnamefont
  {Verde}}, \bibinfo {author} {\bibfnamefont {M.}~\bibnamefont {Dogterom}},
  \bibinfo {author} {\bibfnamefont {E.}~\bibnamefont {Stelzer}}, \bibinfo
  {author} {\bibfnamefont {E.}~\bibnamefont {Karsenti}}, \ and\ \bibinfo
  {author} {\bibfnamefont {S.}~\bibnamefont {Leibler}},\ }\href {\doibase
  10.1083/jcb.118.5.1097} {\bibfield  {journal} {\bibinfo  {journal} {J. Cell
  Biol.}\ }\textbf {\bibinfo {volume} {118}},\ \bibinfo {pages} {1097}
  (\bibinfo {year} {1992})}\BibitemShut {NoStop}%
\bibitem [{\citenamefont {Luchniak}\ \emph {et~al.}(2023)\citenamefont
  {Luchniak}, \citenamefont {Kuo}, \citenamefont {McGuinness}, \citenamefont
  {Sutradhar}, \citenamefont {Orbach}, \citenamefont {Mahamdeh},\ and\
  \citenamefont {Howard}}]{Luchniak2023}%
  \BibitemOpen
  \bibfield  {author} {\bibinfo {author} {\bibfnamefont {A.}~\bibnamefont
  {Luchniak}}, \bibinfo {author} {\bibfnamefont {Y.-W.}\ \bibnamefont {Kuo}},
  \bibinfo {author} {\bibfnamefont {C.}~\bibnamefont {McGuinness}}, \bibinfo
  {author} {\bibfnamefont {S.}~\bibnamefont {Sutradhar}}, \bibinfo {author}
  {\bibfnamefont {R.}~\bibnamefont {Orbach}}, \bibinfo {author} {\bibfnamefont
  {M.}~\bibnamefont {Mahamdeh}}, \ and\ \bibinfo {author} {\bibfnamefont
  {J.}~\bibnamefont {Howard}},\ }\href {\doibase 10.1016/j.bpj.2023.01.020}
  {\bibfield  {journal} {\bibinfo  {journal} {Biophys. J.}\ }\textbf {\bibinfo
  {volume} {122}},\ \bibinfo {pages} {616} (\bibinfo {year}
  {2023})}\BibitemShut {NoStop}%
\bibitem [{\citenamefont {Sept}\ \emph {et~al.}(2003)\citenamefont {Sept},
  \citenamefont {Baker},\ and\ \citenamefont {McCammon}}]{Sept2003}%
  \BibitemOpen
  \bibfield  {author} {\bibinfo {author} {\bibfnamefont {D.}~\bibnamefont
  {Sept}}, \bibinfo {author} {\bibfnamefont {N.~A.}\ \bibnamefont {Baker}}, \
  and\ \bibinfo {author} {\bibfnamefont {J.~A.}\ \bibnamefont {McCammon}},\
  }\href {\doibase doi:10.1110/ps.03187503} {\bibfield  {journal} {\bibinfo
  {journal} {Prot. Sci.}\ }\textbf {\bibinfo {volume} {12}},\ \bibinfo {pages}
  {2257} (\bibinfo {year} {2003})}\BibitemShut {NoStop}%
\bibitem [{\citenamefont {Kononova}\ \emph {et~al.}(2014)\citenamefont
  {Kononova}, \citenamefont {Kholodov}, \citenamefont {Theisen}, \citenamefont
  {Marx}, \citenamefont {Dima}, \citenamefont {Ataullakhanov}, \citenamefont
  {Grishchuk},\ and\ \citenamefont {Barsegov}}]{Kononova2014}%
  \BibitemOpen
  \bibfield  {author} {\bibinfo {author} {\bibfnamefont {O.}~\bibnamefont
  {Kononova}}, \bibinfo {author} {\bibfnamefont {Y.}~\bibnamefont {Kholodov}},
  \bibinfo {author} {\bibfnamefont {K.~E.}\ \bibnamefont {Theisen}}, \bibinfo
  {author} {\bibfnamefont {K.~A.}\ \bibnamefont {Marx}}, \bibinfo {author}
  {\bibfnamefont {R.~I.}\ \bibnamefont {Dima}}, \bibinfo {author}
  {\bibfnamefont {F.~I.}\ \bibnamefont {Ataullakhanov}}, \bibinfo {author}
  {\bibfnamefont {E.~L.}\ \bibnamefont {Grishchuk}}, \ and\ \bibinfo {author}
  {\bibfnamefont {V.}~\bibnamefont {Barsegov}},\ }\href {\doibase
  10.1021/ja506385p} {\bibfield  {journal} {\bibinfo  {journal} {J. Am. Chem.
  Soc.}\ }\textbf {\bibinfo {volume} {136}},\ \bibinfo {pages} {17036}
  (\bibinfo {year} {2014})}\BibitemShut {NoStop}%
\bibitem [{\citenamefont {Hemmat}\ \emph {et~al.}(2019)\citenamefont {Hemmat},
  \citenamefont {Castle}, \citenamefont {Sachs},\ and\ \citenamefont
  {Odde}}]{Hemmat2019}%
  \BibitemOpen
  \bibfield  {author} {\bibinfo {author} {\bibfnamefont {M.}~\bibnamefont
  {Hemmat}}, \bibinfo {author} {\bibfnamefont {B.~T.}\ \bibnamefont {Castle}},
  \bibinfo {author} {\bibfnamefont {J.~N.}\ \bibnamefont {Sachs}}, \ and\
  \bibinfo {author} {\bibfnamefont {D.~J.}\ \bibnamefont {Odde}},\ }\href
  {\doibase 10.1016/j.bpj.2019.08.011} {\bibfield  {journal} {\bibinfo
  {journal} {Biophys. J.}\ }\textbf {\bibinfo {volume} {117}},\ \bibinfo
  {pages} {1234} (\bibinfo {year} {2019})}\BibitemShut {NoStop}%
\bibitem [{\citenamefont {VanBuren}\ \emph {et~al.}(2005)\citenamefont
  {VanBuren}, \citenamefont {Cassimeris},\ and\ \citenamefont
  {Odde}}]{vanBuren2005}%
  \BibitemOpen
  \bibfield  {author} {\bibinfo {author} {\bibfnamefont {V.}~\bibnamefont
  {VanBuren}}, \bibinfo {author} {\bibfnamefont {L.}~\bibnamefont
  {Cassimeris}}, \ and\ \bibinfo {author} {\bibfnamefont {D.~J.}\ \bibnamefont
  {Odde}},\ }\href@noop {} {\bibfield  {journal} {\bibinfo  {journal} {Biophys.
  J.}\ }\textbf {\bibinfo {volume} {89}},\ \bibinfo {pages} {2911} (\bibinfo
  {year} {2005})}\BibitemShut {NoStop}%
\bibitem [{\citenamefont {Coombes}\ \emph {et~al.}(2013)\citenamefont
  {Coombes}, \citenamefont {Yamamoto}, \citenamefont {Kenzie}, \citenamefont
  {Odde},\ and\ \citenamefont {Gardner}}]{Coombes2013}%
  \BibitemOpen
  \bibfield  {author} {\bibinfo {author} {\bibfnamefont {C.}~\bibnamefont
  {Coombes}}, \bibinfo {author} {\bibfnamefont {A.}~\bibnamefont {Yamamoto}},
  \bibinfo {author} {\bibfnamefont {M.}~\bibnamefont {Kenzie}}, \bibinfo
  {author} {\bibfnamefont {D.}~\bibnamefont {Odde}}, \ and\ \bibinfo {author}
  {\bibfnamefont {M.}~\bibnamefont {Gardner}},\ }\href {\doibase
  10.1016/j.cub.2013.05.059} {\bibfield  {journal} {\bibinfo  {journal} {Curr.
  Biol.}\ }\textbf {\bibinfo {volume} {23}},\ \bibinfo {pages} {1342} (\bibinfo
  {year} {2013})}\BibitemShut {NoStop}%
\bibitem [{\citenamefont {Zakharov}\ \emph {et~al.}(2015)\citenamefont
  {Zakharov}, \citenamefont {Gudimchuk}, \citenamefont {Voevodin},
  \citenamefont {Tikhonravov}, \citenamefont {Ataullakhanov},\ and\
  \citenamefont {Grishchuk}}]{Zakharov2015}%
  \BibitemOpen
  \bibfield  {author} {\bibinfo {author} {\bibfnamefont {P.}~\bibnamefont
  {Zakharov}}, \bibinfo {author} {\bibfnamefont {N.}~\bibnamefont {Gudimchuk}},
  \bibinfo {author} {\bibfnamefont {V.}~\bibnamefont {Voevodin}}, \bibinfo
  {author} {\bibfnamefont {A.}~\bibnamefont {Tikhonravov}}, \bibinfo {author}
  {\bibfnamefont {F.}~\bibnamefont {Ataullakhanov}}, \ and\ \bibinfo {author}
  {\bibfnamefont {E.}~\bibnamefont {Grishchuk}},\ }\href {\doibase
  10.1016/j.bpj.2015.10.048} {\bibfield  {journal} {\bibinfo  {journal}
  {Biophys. J.}\ }\textbf {\bibinfo {volume} {109}},\ \bibinfo {pages} {2574}
  (\bibinfo {year} {2015})}\BibitemShut {NoStop}%
\bibitem [{\citenamefont {McIntosh}\ \emph {et~al.}(2018)\citenamefont
  {McIntosh}, \citenamefont {O’Toole}, \citenamefont {Morgan}, \citenamefont
  {Austin}, \citenamefont {Ulyanov}, \citenamefont {Ataullakhanov},\ and\
  \citenamefont {Gudimchuk}}]{McIntosh2018}%
  \BibitemOpen
  \bibfield  {author} {\bibinfo {author} {\bibfnamefont {J.~R.}\ \bibnamefont
  {McIntosh}}, \bibinfo {author} {\bibfnamefont {E.}~\bibnamefont {O’Toole}},
  \bibinfo {author} {\bibfnamefont {G.}~\bibnamefont {Morgan}}, \bibinfo
  {author} {\bibfnamefont {J.}~\bibnamefont {Austin}}, \bibinfo {author}
  {\bibfnamefont {E.}~\bibnamefont {Ulyanov}}, \bibinfo {author} {\bibfnamefont
  {F.}~\bibnamefont {Ataullakhanov}}, \ and\ \bibinfo {author} {\bibfnamefont
  {N.}~\bibnamefont {Gudimchuk}},\ }\href {\doibase 10.1083/jcb.201802138}
  {\bibfield  {journal} {\bibinfo  {journal} {J. Cell Biol.}\ }\textbf
  {\bibinfo {volume} {217}},\ \bibinfo {pages} {2691} (\bibinfo {year}
  {2018})}\BibitemShut {NoStop}%
\bibitem [{\citenamefont {Gudimchuk}\ \emph {et~al.}(2020)\citenamefont
  {Gudimchuk}, \citenamefont {Ulyanov}, \citenamefont {O’Toole},
  \citenamefont {Page}, \citenamefont {Vinogradov}, \citenamefont {Morgan},
  \citenamefont {Li}, \citenamefont {Moore}, \citenamefont {Szczesna},
  \citenamefont {Roll-Mecak}, \citenamefont {Ataullakhanov},\ and\
  \citenamefont {Richard~McIntosh}}]{Gudimchuk2020}%
  \BibitemOpen
  \bibfield  {author} {\bibinfo {author} {\bibfnamefont {N.~B.}\ \bibnamefont
  {Gudimchuk}}, \bibinfo {author} {\bibfnamefont {E.~V.}\ \bibnamefont
  {Ulyanov}}, \bibinfo {author} {\bibfnamefont {E.}~\bibnamefont {O’Toole}},
  \bibinfo {author} {\bibfnamefont {C.~L.}\ \bibnamefont {Page}}, \bibinfo
  {author} {\bibfnamefont {D.~S.}\ \bibnamefont {Vinogradov}}, \bibinfo
  {author} {\bibfnamefont {G.}~\bibnamefont {Morgan}}, \bibinfo {author}
  {\bibfnamefont {G.}~\bibnamefont {Li}}, \bibinfo {author} {\bibfnamefont
  {J.~K.}\ \bibnamefont {Moore}}, \bibinfo {author} {\bibfnamefont
  {E.}~\bibnamefont {Szczesna}}, \bibinfo {author} {\bibfnamefont
  {A.}~\bibnamefont {Roll-Mecak}}, \bibinfo {author} {\bibfnamefont {F.~I.}\
  \bibnamefont {Ataullakhanov}}, \ and\ \bibinfo {author} {\bibfnamefont
  {J.}~\bibnamefont {Richard~McIntosh}},\ }\href {\doibase
  10.1038/s41467-020-17553-2} {\bibfield  {journal} {\bibinfo  {journal} {Nat.
  Commun.}\ }\textbf {\bibinfo {volume} {11}} (\bibinfo {year} {2020}),\
  10.1038/s41467-020-17553-2}\BibitemShut {NoStop}%
\bibitem [{\citenamefont {Ganser}\ and\ \citenamefont
  {Uchihashi}(2019)}]{Ganser2019}%
  \BibitemOpen
  \bibfield  {author} {\bibinfo {author} {\bibfnamefont {C.}~\bibnamefont
  {Ganser}}\ and\ \bibinfo {author} {\bibfnamefont {T.}~\bibnamefont
  {Uchihashi}},\ }\href {\doibase 10.1039/c8nr07392a} {\bibfield  {journal}
  {\bibinfo  {journal} {Nanoscale}\ }\textbf {\bibinfo {volume} {11}},\
  \bibinfo {pages} {125} (\bibinfo {year} {2019})}\BibitemShut {NoStop}%
\bibitem [{\citenamefont {Katsuki}\ \emph {et~al.}(2014)\citenamefont
  {Katsuki}, \citenamefont {Drummond},\ and\ \citenamefont
  {Cross}}]{Katsuki2014}%
  \BibitemOpen
  \bibfield  {author} {\bibinfo {author} {\bibfnamefont {M.}~\bibnamefont
  {Katsuki}}, \bibinfo {author} {\bibfnamefont {D.~R.}\ \bibnamefont
  {Drummond}}, \ and\ \bibinfo {author} {\bibfnamefont {R.~A.}\ \bibnamefont
  {Cross}},\ }\href@noop {} {\bibfield  {journal} {\bibinfo  {journal} {Nat.
  Commun.}\ }\textbf {\bibinfo {volume} {5}},\ \bibinfo {pages} {1} (\bibinfo
  {year} {2014})}\BibitemShut {NoStop}%
\bibitem [{Zab(2026)}]{Zablotsky2026Zenodo}%
  \BibitemOpen
  \href@noop {} {}\bibinfo {howpublished} {Data supplement for "Beyond the Tip:
  Lattice Dynamics, Seams, and the Mechanism of Microtubule Fracture", Zenodo,
  doi:10.5281/zenodo.1742565} (\bibinfo {year} {2026})\BibitemShut {NoStop}%
\end{thebibliography}

%

\end{document}


\title{Beyond the Tip: Lattice Dynamics, Seams, and the Mechanism of Microtubule Fracture}

\author{Amir Zablotsky}
\affiliation{Université Grenoble-Alpes, CNRS, Laboratoire Interdisciplinaire de Physique 38000 Grenoble, France}

\author{Subham Biswas}
\affiliation{Experimental Physics and Center for Biophysics, Saarland University, 66123 Saarbrücken, Germany}

\author{Laura Schaedel}
\email{laura.schaedel@uni-saarland.de}
\affiliation{Experimental Physics and Center for Biophysics, Saarland University, 66123 Saarbrücken, Germany}
\affiliation{PharmaScienceHub (PSH), 66123 Saarbrücken, Germany}

\author{Karin John}
\email{karin.john@univ-grenoble-alpes.fr}
\affiliation{Université Grenoble-Alpes, CNRS, Laboratoire Interdisciplinaire de Physique 38000 Grenoble, France}


\date{\today}

\maketitle

\onecolumngrid

\begin{center}
\vspace*{4ex}
\Large{\textbf{Supplemental Material}}
\end{center}

All bibliographic references included in the Supplemental Material are listed below and also appear in the bibliography section of the main text, though with different numbering.

\renewcommand\thefigure{S\arabic{figure}}    
\setcounter{figure}{0} 

\section{Details of the kinetic Monte Carlo simulations}

The MT lattice is modeled as a square lattice at the scale of the monomer. Each monomer lattice site is identified by a doublet $(i,j)$; each column $i\in (1,\ldots 13)$ in the lattice corresponds to a protofilament. The main seam defined by the lateral contacts between heterotypic units ($\alpha-\beta$, $\beta-\alpha$) is always initially located between protofilaments $i=13$ and $i=1$ at the MT minus end, although this may change due to the presence of monomer defects in the lattice.  The lateral periodic boundary has a three monomers vertical offset to account for the 3-start helix, i.e.~a lattice site at $(1,j)$ forms a lateral contact with a lattice site at $(13,j+3)$. Each lattice site is either vacant or occupied by a monomer, whereby the presence of an $\alpha-$monomer at site $(i,j)$ implies the presence of a $\beta-$monomer at site $(i,j+1)$ (and vice versa) to maintain the dimeric character of tubulin. 
Tubulin dimers detach from the lattice with the rate constant $k_\text{mn}$ (Eq.~(3) in the main text), depending on the number of longitudinal ($m$) and lateral ($n$) monomer neighbors.
The kinetic Monte Carlo process is realized using a rejection free variable step size method \cite{Lukkien1998}.
%
%
Briefly, each simulation step follows the following procedure.
\begin{enumerate}
\item The lattice is scanned and each possible lattice transition (from lattice state $c$ to lattice state $c'$) is tabulated and an effective rate constant $K(c)$ is calculated
\begin{equation}
K(c)=\sum_i k_i n_i(c)\,,
\end{equation} 
whereby $n_i$ denotes the number of occurrences of the lattice transition with rate constant $k_i$.
\item The time increment, $dt$, for the occurrence of the next lattice transition is drawn from an exponential distribution with rate parameter $K(c)$. 
\item Among all possible lattice transitions, a lattice transition with rate constant $k_j$ is randomly chosen with probability 
\begin{equation}
p_j(c)={ n_j(c) k_j\over K(c) }\,.
\end{equation}
\item Among all lattice sites $n_j$ where the lattice transition with rate constant $k_j$ is enabled, a lattice site $l$ is chosen randomly with probability $1\over n_j$.
\item The chosen lattice transition with rate constant $k_j$ at lattice site $l$ is executed, the time is advanced by the increment $dt$ obtained in step 2 and the procedure is repeated starting with step 1.
\end{enumerate}

\section{Discussion of the dependence of the size of damage, $L_\text{f}$, and the time to fracture, $T_\text{f}$, on the lattice anisotropy $A$}

\begin{figure*}[hbt]
\centering
\includegraphics[width=0.8\hsize]{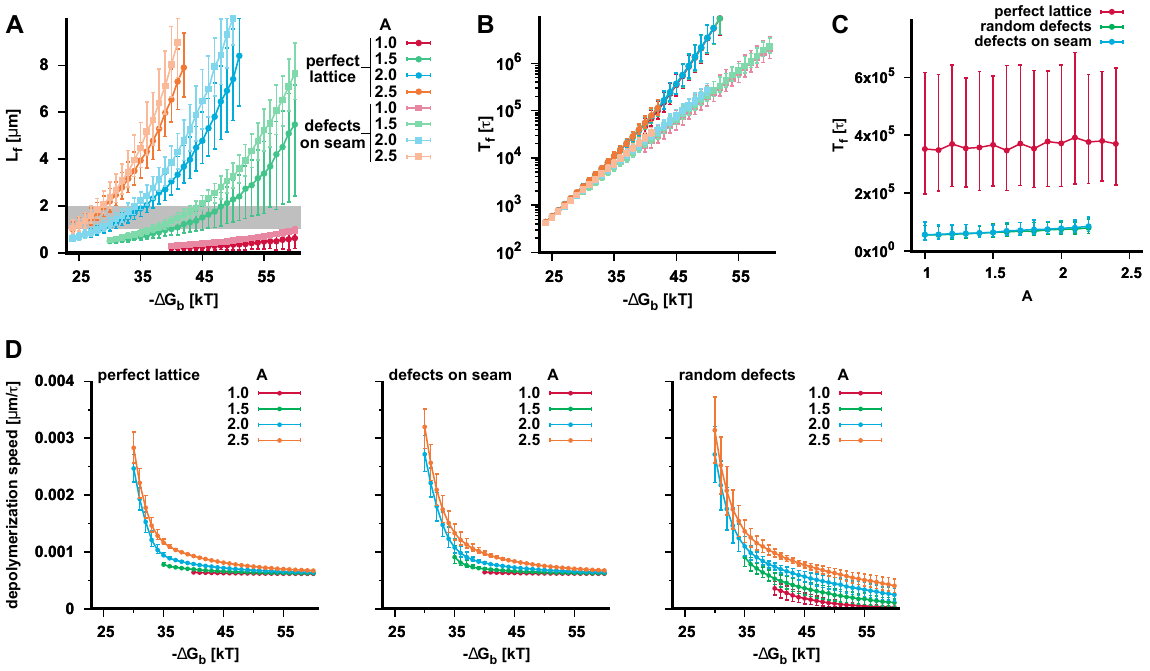}
\caption{Fracture damage size ($L_\text{f}$, A) and time ($T_\text{f}$, B) vs. lattice binding energy ($\Delta G_\text{b}$) for varying anisotropies ($A$). Results for defect-free lattices and seam-localized defects. Gray region in (A): experimental range (see Fig.\,1C).
\textbf{C:} Time to fracture $T_\text{f}$ vs. lattice anisotropy $A$ for constant total lattice binding energy $\Delta G_\text{b}=-45\,$kT. Results for defect-free lattices (perfect lattice), uniformly distributed defects (random defects) and seam-localized defects (defects on seam). Data: median $\pm$ IQR (1000 simulations). The MT length in (A-C) is 10\,$\muup$m.
\textbf{D:} Tip depolymerization speed of 10\,$\muup$m long MTs vs. lattice binding energy ($\Delta G_\text{b}$) for varying anisotropies ($A$). MTs contain no defects (left), seam-localized defects (center), or uniformly distributed defects (right). The defect frequency in (A-D) is 0.2\,$\muup$m$^{-1}$. Symbols and error bars in (A-D) denote the median and IQR of 1000 simulations.}
\label{figS1}
\end{figure*}

In Fig.\,4A,B we evaluated the size of damage $L_\text{f}$ and time to fracture $T_\text{f}$ for 10\,$\muup$m long MTs with a perfect lattice and with a lattice containing monomer vacancies uniformly distributed over the lattice with spatial frequency 0.2\,$\muup$m$^{-1}$. Fig.\,\ref{figS1}A,B shows complementary results for a lattice containing monomers localized on the main seam, which shows the same trends as for the case of uniformly distributed vacancies.  
For a constant lattice energy $\Delta G_\text{b}$ the damage size $L_\text{f}$ increases with anisotropy $A$ (Fig.\,4A). The trend for the time to fracture $T_\text{f}$ is less obvious  in Fig.\,4B, but it becomes clearer in Fig.\,\ref{figS1}C. For a perfect lattice at lattice binding energy $\Delta G_\text{b}=-45$\,kT the time to fracture is independent of anisotropy $A$. For a lattice with defects, the time to fracture increases weakly with anisotropy $A$.

In a perfect lattice, the time to fracture is the sum of two components: the time required to remove the first dimer from the fully occupied lattice, $T_1$, and the time for the damage to propagate across all protofilaments, $T_2$. $T_1$ is given by (cf.\,Eq.\,(3) in the main text)
\begin{equation}
T_1={1\over N k_{24}} =  {\tau \over N} e^{-{\beta \over 2} \Delta G_\text{b}}
\end{equation} 
with $N$ denoting the total number of dimers contained in the lattice. 
For 10\,$\muup$m long MTs with lattice energy $-\Delta G_\text{b}>40$\,kT, $T_1$ dominates the time to fracture, $T_\text{f}$. This is evident from the marked increase in $T_\text{f}$ for a perfect lattice compared to a lattice containing defects (see Fig.\,4B and Fig.\,\ref{figS1}C). In the perfect lattice, any influence of lattice anisotropy on $T_\text{f}$ is therefore masked by the influence of $T_1$.

For a lattice containing defects, the time to fracture, $T_\text{f}$, equals $T_2$. As mentioned in the main text, the lateral front propagation speed of the damage, $v_\text{lat} \sim N_\text{long} k_{22}$, depends on two factors: the off-rate constant $k_{22}$ of dimers with two longitudinal and one full lateral neighbor (blue dimers in Fig.\,3A-C), and the number of dimers lining the lateral boundaries of the damaged region, $N_\text{long}$. Increasing lattice anisotropy slows down lateral fracture propagation by reducing the dimer off-rate constant (recall that $\Delta G_\text{long}<0$)
\begin{equation}
k_{22}={1\over \tau} e^{\beta \left(2\Delta G_\text{long}+\Delta G_\text{lat}-{\Delta G_\text{b}\over 2} \right)}={1\over \tau} e^{\beta\Delta G_\text{long}}\,.
\end{equation}
However, $v_\text{lat}$ increases with $N_\text{long}$, which is determined by the longitudinal front speed $v_\text{long}$. $v_\text{long}$ depends on the structure of the longitudinal front (lateral extension, presence of seams). This coupling between lateral and longitudinal front propagation indicates that damage propagation is not a trivial process.

For the sake of argument and as an order of magnitude estimate, we assume that the longitudinal front speed is driven by the presence of a single seam dimer (red in Fig.\,3A-C), i.e., $v_\text{long}\sim{1\over 2}k_{13}$. Here,  the factor of ${1\over 2}$ arises because the removal of one seam dimer advances the front by one monomer. Solving for the longitudinal and lateral size of the damage, $N_\text{long}$ and $N_\text{lat}$, respectively -- both of which spread with speed $2 v_\text{long}$ and $2 v_\text{lat}$ -- we find, after some manipulation
\begin{eqnarray}
N_\text{long} (t)& = & k_{13} t \\
N_\text{lat} (t)& = & k_{13}k_{22} t^2\,.
\end{eqnarray} 
The time to fracture is given by $N_\text{lat}(T_\text{f})=N_{PF}$ with $N_\text{PF}=13$ denoting the number of protofilaments in the lattice
\begin{equation}
T_\text{f}=\sqrt{N_\text{PF}\over k_{13} k_{22}}\,.
\end{equation}
The damage size at fracture in numbers of dimers, $N_\text{long}(T_\text{f})$, is given by
\begin{equation}
N_\text{long}(T_\text{f})=k_{13}T_\text{f}=\sqrt{N_\text{PF}{k_{13} \over k_{22}}}\,.
\end{equation}
Using Eq.\,(3) of the main text we find
\begin{eqnarray}
T_\text{f} & = & \tau\sqrt{N_\text{PF}} e^{-{\beta \over 2}\left(\Delta G_\text{long}+{1\over 2}\Delta G_\text{lat}\right)}=\tau\sqrt{N_\text{PF}} e^{-\left({\beta \over 4} {A+{1\over 2}\over A+1}\Delta G_\text{b}\right)}\,. \label{Tf}\\
N_\text{long}(T_\text{f}) & = & \sqrt{N_\text{PF}} e^{-{\beta \over 2}\left(\Delta G_\text{long}-{1\over 2}\Delta G_\text{lat}\right)}=\sqrt{N_\text{PF}} e^{-\left({\beta \over 4}{A-{1\over 2}\over A+1}\Delta G_\text{b}\right) }\,. \label{Nlongf}
\end{eqnarray}
Eqs.\,\eqref{Tf} and \eqref{Nlongf} show that $T_\text{f}$ and $ N_\text{long}(T_\text{f})$ increase with anisotropy $A$ for constant $\Delta G_\text{b}$ (note that $\Delta G_\text{b}, \Delta G_\text{long}, \Delta G_\text{lat}<0$). However, for $A>1$ this increase is much stronger for the damage size, $N_\text{long}(T_\text{f})$, than for the fracture time $T_\text{f}$.

\section{Calibration of the time scale $\tau$}

 \label{sec-timescale}

We calibrate the timescale of our simulations ($\tau$ in Eq.\,(3)) by simulating MT depolymerization from the plus-tip and calculating the depolymerization speed $\hat{v}$ in units of $\muup$m/$\tau$. The simulated tip speed $\hat{v}$ is then used to calculate the time scale $\tau$ from the experimentally measured depolymerization speeds $v=10\ldots30\,\muup$m/min \cite{Mitchison1984,Cassimeris1986,Walker1988,Verde1992,Luchniak2023} using the relation
\begin{equation}
\tau={\hat{v}\over v}\,.
\end{equation}
Fig.\,\ref{figS1}D shows exemplarily the dependence of $\hat{v}$ on the binding energy $\Delta G_\text{b}$ for various values of the lattice anisotropy $A$ for three different MT lattice types (perfect, with defects localized on the main seam or with uniformly distributed defects, defect frequency 0.2\,$\muup$m$^{-1}$).
For the perfect lattice and for a lattice with only on-seam defects, the trends for the depolymerization speed are similar.
For higher lattice energies $-\Delta G_\text{b}\gtrsim 40$\,kT the tip depolymerization approaches the limit of helical depolymerization $\hat{v}=1\,\muup$m$/(N\tau)\approx 6.15\times 10^{-4}\,\muup$m/$\tau$, with $N$ denoting the number of dimers in a MT of length 1\,$\muup$m, i.e.~$N=125\times 13$.
For lower lattice energies $-\Delta G_\text{b}\lesssim 40$\,kT and high anisotropies the presence of defects slightly accelerates depolymerization, since MTs are already losing dimers in the vicinity of defects during tip depolymerization. This effect is dependent on the MT length and frequency of defects used for the simulations.   
For randomly distributed defects at higher lattice energies and low anisotropies, the presence of off-seam defects and multiseam structures considerably slows down the depolymerization speed compared to the perfect lattice. On-seam dimers at the main seam do not influence tip depolymerization speed, as they only cause a lateral shift of the main seam without altering the lattice structure.

\section{Influence of the defect frequency on the damage size, $L_\text{f}$, and time to fracture, $T_\text{f}$}

 \begin{figure}[hbt]
\centering
\includegraphics[width=0.5\hsize]{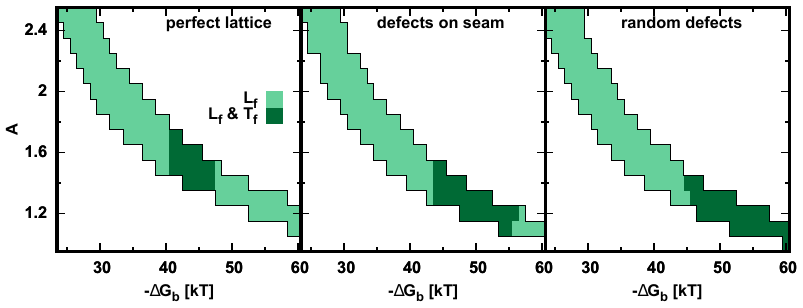}
\caption{Lattice parameters $\left(\Delta G_b,A\right)$ that reproduce experiments assuming 10\,$\muup$m long MTs with a perfect lattice (perfect lattice), containing defects localized on the main seam (defects on seam), or uniformly distributed defects (random defects).
Light green: parameters reproducing $L_\text{f}$; dark green: parameters reproducing both $L_\text{f}$ and $T_\text{f}$.
The spatial frequency of defects is 0.1\,$\muup$m$^{-1}$.
}
\label{figS2}
\end{figure}

To test the robustness of our simulations of MT fracture shown in Fig.\,4 with a spatial frequency of monomer defects of 0.2\,$\muup$m$^{-1}$ we performed an identical set of simulations for 10\,$\muup$m long MTs with a defect frequency of 0.1\,$\muup$m$^{-1}$, i.e. we determined the damage size at fracture and the time to fracture from 1000 independent simulations. Furthermore we estimated the time scale $\tau$ from tip depolymerization simulations for MTs containing the same defect distributions as in the fracture experiments.
Fig.\,\ref{figS2} shows the range of lattice parameters compatible with fracture experiments (Fig.\,1B,C) for the perfect lattice or for a lattice containing defects (spatial frequency 0.1\,$\muup$m$^{-1}$) either localized on the seam (defects on seam) or distributed uniformly throughout the lattice (random defects).
The trend is the same as for MTs with a higher spatial frequency (0.2\,$\muup$m$^{-1}$; Fig.\,4C). The presence of defects implies higher binding energy $\Delta G_\text{b}$ and a lower anisotropy $A$.

\section{Estimation of GDP-lattice parameters from the literature}

In general, the free energy of binding of a dimer to the lattice contains an enthalpic contribution, determined by the potentials of the interacting side chains ($\Delta H$) and two entropic contributions due to the loss of conformational degrees of freedom ($-T\Delta S_\text{side chain}$) and the loss of rigid body rotational and translational degrees of freedom of the dimer upon integrating the lattice ($-T\Delta S_\text{rigid body}$) \cite{Hemmat2019}. Here, in our kinetic Monte Carlo model, we have used a description where the binding energy of a dimer is given by the sum of contributions from lateral and longitudinal nearest neighbor contacts, whereby each individual contact between two neighboring dimers contains the enthalpic part $\Delta H$ and the atomistic entropic part $-T\Delta S_\text{side chain}$. The rigid-body contribution of the entropy $-T\Delta S_\text{rigid body}$ is resorbed into the off-rate constant  $\tau^{-1}$ and does not appear explicitly in our description.

For comparison with lattice parameters in the literature obtained from kinetic Monte Carlo modeling of the tip dynamics, it is important to note that the rigid body entropic loss is typically contained in the longitudinal binding energy, i.e. $\Delta G^\ast_\text{long}=\Delta H_\text{long}-T(\Delta S_\text{side chain,long}+\Delta S_\text{rigid body})$. The lateral binding energy contains only the atomistic entropic contribution as in our model $\Delta G_\text{lat}=\Delta H_\text{lat}-T\Delta S_\text{side chain,lat}$. 
The use of $\Delta G^\ast_\text{long}$ in tip growth models has practical reasons, since the binding/unbinding of dimers to and from the microtubule tip always involves a single longitudinal bond and a variable number of lateral bonds and the free energy of polymerization is calculated from the principle of detailed balance from the dimer on- and off-rate constants. In our simulations, dimers detach from a variable number of longitudinal and lateral neighbors, which makes using $\Delta G^\ast_\text{long}$ impractical.  

The loss in entropy due to rigid body motions has been estimated in Ref.~\cite{Castle2013} to be $\Delta G_\text{S,rigid body}=10\,$kT and is consistent with previous estimations of 10-20\,kT \cite{Erickson1989}. Assuming the lower estimate from Ref.~\cite{Castle2013}, $\Delta G_\text{S,rigid body}=10\,$kT, we can establish the link of the longitudinal binding energy $\Delta G_\text{long}$ used in our model  and the longitudinal binding energy $\Delta G^\ast_\text{long}$ used in tip growth models as $\Delta G_\text{long}\approx\Delta G^\ast_\text{long}-10\,$kT.

Ref.~\cite{Hemmat2019} combines detailed molecular dynamics, coarse grained brownian dynamics and kinetic simulations to a very detailed analysis of the energetics of the MT tip lattice giving rise to the following numbers (see Table 5 in Ref.~\cite{Hemmat2019}).
A single longitudinal contact has an enthalpic contribution $\Delta H_\text{long} =-17.5$\,kT and an entropic contribution $\Delta G_\text{S}$ (rigid-body and atomistic) of 11.1\,kT. A longitudinal and lateral contact (GDP) has an enthalpic contribution $\Delta H=-29.7\,$kT with a total entropic contribution of $\Delta G_\text{S}=18.6\,$kT.
The increase in $\Delta G_\text{S}$ from one longitudinal neighbor to one longitudinal and one lateral neighbors stems mainly from the atomistic entropy of the lateral tubulin-tubulin contact, i.e.~$\Delta G_\text{S, side chain, lat}=[18.6-11.1]$\,kT=7.5\,kT. Assuming that each interface (lateral monomer-monomer, longitudinal tubulin-tubulin) has approximately the same atomistic entropy contribution we find a rigid-body entropy contribution of 
$\Delta G_\text{S,rigid body}=[11.1-3.75]$\,kT=7.35\,kT.
Integrating now the enthalpic and atomistic entropy contributions into a binding free energy, we find for the longitudinal binding energy $\Delta G_\text{long}=[-17.5+3.75]\,kT=-13.75$\,kT and for the lateral binding energy $\Delta G_\text{lat}=[(-29.7+17.5)+7.5]\,$kT=-4.7\,kT. This results in the total lattice energy and anisotropy in our simulations of $\Delta G_\text{b}=-36.9$\,kT and $A=2.9$.

%